\documentclass[aps,prd,superscriptaddress,nofootinbib]{revtex4-1}

\usepackage{textcomp} 
\usepackage{latexsym,cancel,amssymb,amsmath,verbatim,mathrsfs}
\usepackage{graphicx}
\usepackage[colorlinks,citecolor=blue]{hyperref}
\usepackage{subcaption}
\usepackage{mathtools}

\newcommand{\beq}{\begin{equation}}
\newcommand{\eeq}{\end{equation}}
\newcommand{\bea}{\begin{eqnarray}}
\newcommand{\eea}{\end{eqnarray}}
\newcommand{\met}{\not{\!\!{\rm E}}_{T}}
\newcommand{\nn}{\nonumber}
\newcommand{\sw}{s_W}
\newcommand{\cw}{c_W}
\newcommand{\eemm}{e^+e^- \to \mu^+\mu^-}
\newcommand{\pslash}[1]{\not{\!#1}}

\allowdisplaybreaks[1]

\begin{document}

\title{Probing dark particles indirectly at the CEPC}

\author{Qing-Hong Cao}
\email{qinghongcao@pku.edu.cn}
\affiliation{Department of Physics and State Key Laboratory of Nuclear Physics and Technology, Peking University, Beijing 100871, China}
\affiliation{Collaborative Innovation Center of Quantum Matter, Beijing 100871, China}
\affiliation{Center for High Energy Physics, Peking University, Beijing 100871, China}

\author{Yang Li}
\email{johnpaul@pku.edu.cn}
\affiliation{Department of Physics and State Key Laboratory of Nuclear Physics and Technology, Peking University, Beijing 100871, China}

\author{Bin Yan}
\email{binyan@pku.edu.cn}
\affiliation{Department of Physics and State Key Laboratory of Nuclear Physics and Technology, Peking University, Beijing 100871, China}

\author{Ya Zhang}
\email{zhangya1221@pku.edu.cn}
\affiliation{Department of Physics and State Key Laboratory of Nuclear Physics and Technology, Peking University, Beijing 100871, China}

\author{Zhen Zhang}
\email{zh.zhang@pku.edu.cn}
\affiliation{Center for High Energy Physics, Peking University, Beijing 100871, China}

\begin{abstract}
When dark matter candidate and its parent particles are nearly degenerate, it would be difficult to probe them at the Large Hadron Collider directly. We propose to explore their quantum loop effects at the CEPC through the golden channel process $e^+e^-\to \mu^+\mu^-$. We use a renormalizable toy model consisting of a new scalar and a fermion to describe new physics beyond the Standard Model. The new scalar and fermion are general multiplets of the $SU(2)_L\times U(1)_Y$ symmetry, and couple to the muon lepton through Yukawa interaction. We calculate their loop contributions to anomalous $\gamma\mu^+\mu^-$ and $Z\mu^+\mu^-$  couplings which can be applied to many new physics models. The prospects of their effects at the CEPC are also examined assuming a 2\textperthousand~accuracy in the cross section measurement. 
\end{abstract}

\maketitle

\section{Introduction}

One of the major tasks of particle physics is to understand the particle nature of dark matter~\cite{Ahmed:2009zw,Bertone:2004pz,Akerib:2013tjd,Langacker:2008yv}. As the dark matter candidate does not register at the detector and induce a large missing transverse momentum ($\met$), one usually searches for the dark matter candidate in the signature of a large $\met$ together with a bunch of visible particles in the standard model (SM). The method is valid only when there is a large mass gap between the dark matter candidate and its parent particle. 
However, there could be a scenario in which the dark matter candidate ($Y$) and its parent particle ($X$) are nearly degenerate, e.g. $X\to Y+a$, where $a$ denotes the SM particles. The energy of $a$ tends to 0 ($E_a\to 0$) in the degenerate limit of $X$ and $Y$. The particle $a$'s (or their decay products if $a$'s are not stable) are very soft and cannot register in the detector. It is hard to directly observe or test such new physics signals at the Large Hadron Collider (LHC), and we name it as a ``nightmare" scenario.

On the other hand, the new physics particles affect the SM processes through quantum loop corrections, no matter whether they are degenerate or not. Such quantum corrections, if large enough, could be detected at the electron-positron colliders, e.g. the Circular electron-positron collider (CEPC), FCC-ee or International Linear Collider (ILC).  In this work we focus on the ``nightmare" scenario and explore the potential of measuring the new physics effects in the scattering of $\eemm$ at the CEPC with a center of mass energy of 240~GeV. The $\eemm$ channel is known as the golden channel which serves as a precision candle owing to its clean background and high detection efficiency~\cite{Agashe:2014kda}.  A relative precision of 2\textperthousand~on $\sigma(\eemm)$ can be reached at ILC~\cite{Riemann:2001bb,Baer:2013cma}, and the CEPC~\cite{CEPC-SPPCStudyGroup:2015csa} is expected to achieve a comparable accuracy.

Dark scalars appear often in various new physics models and have been studied extensively in the literature~\cite{Jungman:1995df,Fox:2008kb,Cao:2009yy,Bertone:2004pz,Agrawal:2014ufa}. Rather than considering a specific complete model, we use a simple toy model to describe the new physics beyond the SM. The toy model consists of a new complex scalar multiplet ($S$) and a vector-like fermion ($F$). We demand that the neutral component of $S$ serve as the dark matter candidate, while the fermion $F$ facilitates the Yukawa coupling of $S$ to $\mu^-$. In practice we require that $F$ be slightly heavier than $S$ such that it can decay into $S$ and muon lepton pairs~\footnote{Note that the vector-like fermion $F$, except for a weak gauge singlet, cannot play the role of dark matter candidate as it is constrained severely by the direct detection of the dark matter. However, for the scalar dark matter, it is easy to escape the constraint from LUX data~\cite{Akerib:2015rjg} if a small mass splitting is generated between the real and imaginary components of the neutral complex scalar.}.
Our toy model respects the SM gauge symmetry $SU(3)_C\times SU(2)_L\times U(1)_Y$ and is renormalizable. Therefore, it can be viewed as a simplified version of a UV-completion model and can be generalized to many new physics models, e.g., the lepto-philic dark matter models~\cite{Chen:2008dh,Yin:2008bs,Fox:2008kb,Bi:2009md,Cao:2009yy,Cao:2014cda}. 
To ensure the stability of the dark matter candidate, we restrict the mixing of such exotic particles with the SM particles through an exact $Z_2$ symmetry, under which the SM fields are all even, whereas the new fields are odd. As a result, the SM particles can only interact with a pair of those exotic particles at a time. 

We emphasize that the new physics particles in our toy model can be light, say around $\mathcal{O}(100~{\rm GeV})$, such that the approach of effective field theory~\cite{Hagiwara:1986vm,Falkowski:2015krw,Ellis:2015sca,Wells:2015eba,Bian:2015zha,Cao:2007fy,Cao:2009uw} no longer works, and the full one loop calculation is necessary to address its effects. We use the dimensional regularization to calculate the loop corrections in the on-shell renormalization scheme~\cite{Denner:1991kt,Aoki:1982ed}. The analytical results are written in terms of the Passarino-Veltman scalar functions~\cite{Passarino:1978jh,tHooft:1978xw}. 

The paper is organized as follows. In Sec.~II we first introduce our simplified new physics model with new dark scalar and fermion multiplets. We then calculate the anomalous $\gamma\mu^+\mu^-$ and $Z\mu^+\mu^-$ couplings in the on-shell renormalization scheme. A simple form of those anomalous couplings are also derived in the approximation of large mass expansion. In Sec.~III we evaluate the numerical effects of those anomalous couplings on the cross section of $\eemm$. After taking into account the constraints from dark matter searches at the LHC, we discuss the potential of measuring the loop effects of those dark scalars and fermions through the $\eemm$ channel at the CEPC. Finally, we conclude in Sec.~IV.

\section{Anomalous couplings of $\gamma\mu^+\mu^-$ and $Z\mu^+\mu^-$}

We calculate the loop correction to the scattering of $e^-(p_-) e^+(p_+) \to \mu^-(k_-)\mu^+(k_+)$ from a vector-like fermion $F$ and a scalar $S$, where $p_{\pm}$ and $k_{\pm}$ are the momenta of the electrons and muons.  The new fermion and scalar couple to the SM particles through the following interaction:
\beq\label{mumulag}
\Delta{\cal L}= \bar{F}(iD\!\!\!\!/-M_F)F+|D_\nu S|^2-M_S^2 S^\dagger S -V(S,H)+ {\cal L}_{\rm Yuk},
\eeq
where $D_{\nu,i}=\partial_\nu-igW^a_\nu T^a_i -ig^\prime B_\nu Y_i$ is the usual covariant derivative with $T^{1,2,3}_i$ and $Y_i$ being the  $SU(2)_L$ and $U(1)_Y$ generators of the $i$ field ($i=F,~S$), respectively, and $g$ and $g^\prime$ being the corresponding coupling strengths. $W_\nu^a$ and $B_\nu$ are the weak eigenstate gauge fields, which are related to the weak bosons by $Z_\nu=\cw W^3_\nu-\sw B_\nu$, $A_\nu= \sw W^3_\nu+\cw B_\nu $ and $W^\pm_\nu=(W^1_\nu \mp iW^2_\nu)/\sqrt 2$, where $\sw=\sin\theta_W,~\cw=\cos\theta_W$, with $\theta_W$ being the weak mixing angle. $V(S,H)$ denotes a general scalar potential.
${\cal L}_{\rm Yuk}$ denotes the Yukawa interaction of $F$, $S$ and $\mu^-$; depending on the weak isospins of the $F$ and $S$ fields ($I_F$ and $I_S$), they may couple to either the SM left-handed doublet $\mu_L$ when $I_F=I_S\pm1/2$, or the right-handed singlet $\mu_R$ when $I_F=I_S$.  Besides, the gauge interaction in the first two terms in Eq.~\ref{mumulag} also enters into the loop corrections. We assume no Yukawa interaction of the electron with the new physics fields $F$ and $S$, and ignore the electron mass in our calculations. We shall elaborate the anomalous couplings induced by the Yukawa interaction and the purely gauge interaction separately. 

The demand that $S$ contain an electrically neutral component as the dark matter candidate restricts the value of $Y_S$ as follows,
\begin{equation}\label{modelregion}
    Y_S\in\Big\{-I_S,-I_S+1,\dotsc,I_S-1,I_S\Big\}.
\end{equation}
In this section we first calculate the anomalous couplings of $\gamma\mu^+\mu^-$ and $Z\mu^+\mu^-$ for generic $I_S$ and $Y_S$. The analytical results of our simplified model are for arbitrary representations of $F$ and $S$, and they can be applied to many new physics models. The requirement of having the dark matter component in $S$ is taken into account in our numerical discussion given in Sec.~III.

\subsection{Anomalous couplings induced by the Yukawa interaction}
\subsubsection{\it The $S\bar{\mu}_LF$ coupling scenario}
\vspace*{3mm}

\begin{figure}
    \begin{subfigure}[b]{0.22\textwidth}
        \includegraphics[width=\textwidth]{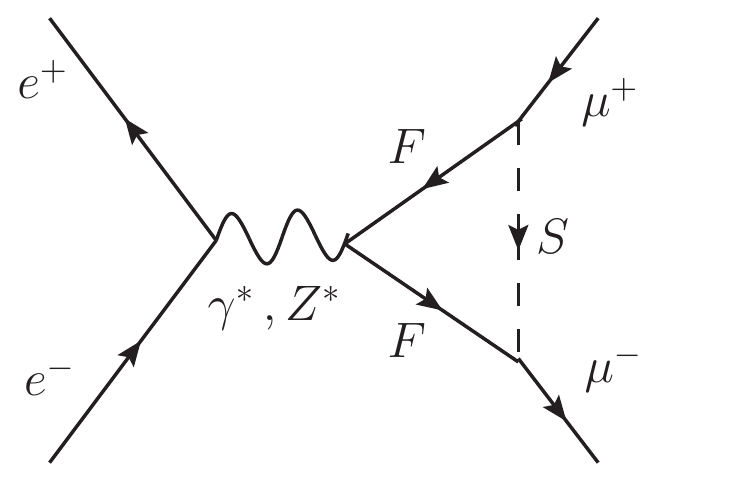}
        \subcaption{}\label{mumu: loop-f-s-F}
    \end{subfigure}
    \begin{subfigure}[b]{0.22\textwidth}
        \includegraphics[width=\textwidth]{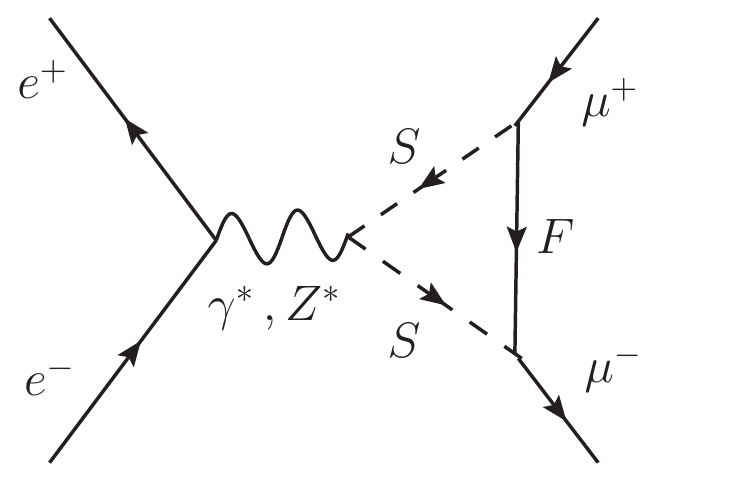}
        \subcaption{}\label{mumu: loop-f-s-S}
    \end{subfigure}
    \begin{subfigure}[b]{0.22\textwidth}
        \includegraphics[width=\textwidth]{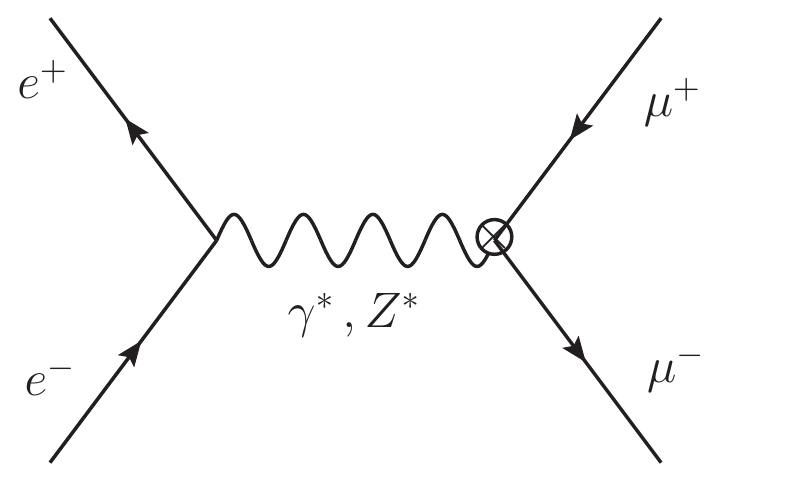}
        \subcaption{}\label{mumu: ct-f-s}
    \end{subfigure}    
    \begin{subfigure}[b]{0.22\textwidth}
        \includegraphics[width=\textwidth]{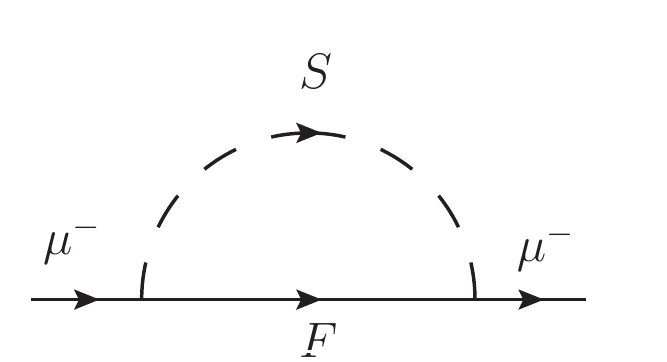}
        \subcaption{}\label{mumu: se-mumu}
    \end{subfigure}
    \caption{Feynman diagrams of Yukawa corrections (a,~b,~c) and the muon self-energy diagram (d).}
    \label{fig:mumu}
\end{figure}

When $I_F=I_S\pm1/2$, $F$ and $S$ couple to the SM left-handed doublet $\mu_L$ through the following Yukawa interaction,
\beq\label{yukl}
{\cal L}_{\rm Yuk}= ~y C_{ijk} S^i {\bar{\mu}_L}^k F^j + h.c.~,
\eeq
where $y$ is the coupling strength and $C_{ijk}=\langle I_{\mu_L} k|I_SI_F;ij\rangle$ are the Clebsch-Gordan (CG) coefficients to render $\Delta{\cal L}$ invariant under the $SU(2)_L$ gauge group. The indices $i,j,k$ label the $T^3$ components of the $S$, $F$ and $\mu_L$ fields, respectively.
At one-loop level, the $\eemm$ process receives corrections from the diagrams in Fig.~\ref{fig:mumu}. Notice that the Yukawa interaction only enters into the self-energy correction of the muon $\mu^-$, but does not enter into the self-energy corrections of the weak gauge bosons. Therefore, it does not renormalize the weak sector.

We parameterize the loop corrections to the $CP$ conserving anomalous couplings of $V\mu^+\mu^- $ with $V=\gamma,Z$ as following~\cite{Stange:1993td,PhysRevD.79.015004}
\beq\label{eqn:mmlorentz}
-i e  \bar{u}\left(k_-\right) \left(\alpha _V\gamma ^{\nu }+\text{i$\beta $}_V  \sigma ^{\nu \rho }q_{\rho }+ \xi _{1,V}\gamma ^{\nu }\gamma _5 +  \xi _{2,V}q^{\nu }\gamma _5\right)v\left(k_+\right),
\eeq
where $e$ is the electrical coupling strength and $q=k_-+k_+$. Among the four interaction terms, only the vector and axial vertices $\gamma^\mu$ and $\gamma^\mu\gamma_5$ are renormalized by the vertex counterterms. The remaining loop-induced Lorentz structures are ultra-violet (UV) finite by themselves, therefore, we decompose Eq.~\ref{eqn:mmlorentz} by,
\begin{align}\label{ffdecomp}
\alpha_{V} &=\alpha_{V,\triangle}+\delta \alpha_{V},\quad
\beta_{V} =\beta_{V,\triangle},\quad
\xi_{1,V} =\xi_{1,V,\triangle}+\delta \xi_{1,V},\quad
\xi_{2,V} =\xi_{2,V,\triangle},
\end{align}
where the couplings with subscriptions $\triangle$ denote the contributions from the triangle loop corrections, and the $\delta \alpha_{V}, \delta \xi_{1,V}$ terms represent the contributions from the vertex counterterms, as depicted in Fig.~\ref{mumu: ct-f-s}. They are given by 
\begin{align}
\delta \alpha_{\gamma} &=-\frac{1}{2} Q \left({\delta Z}_{\mu }^R+{\delta Z}_{\mu }^L\right),&&
\delta \alpha_{Z}=\frac{1}{2} \left(g_R{\delta Z}_{\mu }^R+g_L{\delta Z}_{\mu }^L\right),\nn\\  
\delta \xi_{1,\gamma} &=-\frac{1}{2} Q \left({\delta Z}_{\mu }^R-{\delta Z}_{\mu }^L\right),&&  
\delta \xi_{1,Z}=\frac{1}{2} \left(g_R{\delta Z}_{\mu }^R-g_L{\delta Z}_{\mu }^L\right),
\end{align}
where $\delta Z_\mu^{L/R}$ are the wave function renormalization constants of  $\mu^-_{L/R}$, and
\beq
g_L=\frac{T^3-s_W^2Q }{-c_W s_W}, \qquad g_R=\frac{-s_W^2Q}{-c_W s_W},
\eeq
with $Q=-1$ and $T^3=-1/2$ being the electroweak quantum numbers of $\mu_L^-$.
The renormalization constants $\delta Z_\mu^{L/R}$'s are determined from the muon self-energy corrections (see Fig.~\ref{mumu: se-mumu}), 
\begin{equation}
\Sigma\left(\pslash{p}\right)=\pslash{p}\big[\Sigma_{L}\left(p^{2}\right)P_L+\Sigma_{R}\left(p^{2}\right)P_R\big]+m_\mu \Sigma_{S}\left(p^{2}\right),
\end{equation}
where $P_{L/R}$ are the left/right-handed chirality projectors and $m_\mu$ is the muon mass.
In the on-shell scheme, the finite parts of the counterterms are determined by the requirement that the residue of the fermion propagator at the mass pole is equal to one~\cite{Denner:1991kt,Aoki:1982ed}. Therefore, the wave function renormalization constants are fixed by,
\beq
\delta Z_{\mu}^{L,R} = - m_{\mu}^{2}\frac{\partial}{\partial p^{2}} {\Re}\,\left[\left.\Sigma_{L}(p^{2}) +\Sigma_{R}(p^{2}) +2\Sigma_{S}(p^{2})\right]\right\vert_{p^{2}=m_{\mu}^{2}}-{\Re}\,\Sigma _{L,R}(m_{\mu}^{2}).
\eeq
where $\Re$ denotes taking the real part.

Now we turn to the triangle loop contributions. We first evaluate the $W^3\mu^+\mu^-$ and $B\mu^+\mu^-$ triangle integrals, and derive the $\gamma\mu^+\mu^-$ and $Z\mu^+\mu^-$ vertices using the defining relations $Z_\nu=\cw W^3_\nu-\sw B_\nu$ and $A_\nu= \sw W^3_\nu+\cw B_\nu $. Taking the $W^3\mu^+\mu^-$ loop diagram in Fig.~\ref{mumu: loop-f-s-F} as an example, upon summing over the loop particle components $(S^i,F^j,F^k)$, it is factorized into a generic one-loop integral, multiplied by
\begin{align}
J_{F 3}&=\sum\limits_{ijk}C_{ik-\frac{1}{2}}T_{F,kj}^3C_{ij-\frac{1}{2}}^{*}=\sum\limits_{ijk}\left< \frac{1}{2}-\frac{1}{2} \bigg|I_SI_F;ij\right>\left<I_Fk\Big| \hat J_F^3 \Big|I_Fj\right>\left< I_SI_F;ij\bigg|\frac{1}{2}-\frac{1}{2}\right>\equiv\left< \frac{1}{2}{-\frac{1}{2}}\bigg|\hat J_{F}^3\bigg|\frac{1}{2}{-\frac{1}{2}}\right>,
\end{align}
where $\hat J_F^{3}$ is the third angular momentum operator of the $F$ field. The $B\mu^+\mu^-$ loop is obtained by substituting $T_{F,kj}^3$ with $Y_F\delta_{kj}$ in the formula above, yielding simply $Y_F$. The evaluation of the triangle loop diagram in Fig.~\ref{mumu: loop-f-s-S} is similar, giving $J_{S3}$ and $Y_S$ as group factors. Note that $\hat J_F+\hat J_S=\hat J_{\mu_L}$, we thus have the relation $J_{S3}+J_{F3}=J_{\mu_L3}=T^3_{\mu_L^-}=-1/2$. We also have $Y_S+Y_F=Y_{\mu_{ L}}=-1/2$ due to the $U(1)_Y$ invariance.
Here we choose $J_{S3}$ and $Y_S$ as the independent model parameters, and $J_{S3}$ is worked out to be
\begin{equation}\label{GF:JS3}
    J_{S3}=
    \begin{cases}
      \dfrac{1}{3}I_S, & \quad \text{for  }I_F=I_S+\dfrac{1}{2},\\[3mm]
      -\dfrac{1}{3}(I_S+1), & \quad \text{for  }I_F=I_S-\dfrac{1}{2}.
    \end{cases}
\end{equation}
The generic one-loop triangle integrals are evaluated by reducing them fully into the $B_0$ and $C_0$ scalar functions~\cite{Passarino:1978jh,tHooft:1978xw}. After summing the triangle loop contributions with the counterterms according to Eq.~\ref{ffdecomp}, we obtain the full results in terms of scalar functions, which are listed in App.~\ref{app:yukawaleft}. To manifest the cancellation of the UV-divergences, and also to show the decoupling effect explicitly when the loop particles mass $M_F=M_S=M$ is large, we derive those anomalous couplings in the approximation of large mass expansion. See App.~\ref{app:lme} for the approximate expressions of the $B_0$ and $C_0$ scalar functions. The results are given as follows, 
\begin{align}\label{ancp:mumuleft}
\alpha_{\gamma } &=+\frac{\left| y\right| ^2}{768 \pi ^2}\frac{s}{M^2}\left(2J_{S3}+2 Y_S+3\right),&
\alpha _Z& =  +\frac{\left| y\right| ^2}{768 \pi ^2 c_W s_W}\frac{s}{M^2}\Big(2 c_W^2 J_{S3}-2 s_W^2 Y_S-3 s_W^2+\frac{3}{2}+\frac{ m_{\mu }^2}{s}\Big),\nn\\
\beta _{\gamma }&= +\frac{\left| y\right| ^2}{768 \pi ^2}\frac{m_{\mu }}{M^2}\left(4J_{S3}+4 Y_S+2\right),&
\beta _Z &= +\frac{\left| y\right| ^2}{768 \pi ^2 c_W s_W}\frac{m_{\mu }}{M^2}\Big(4 c_W^2 J_{S3}-4 s_W^2 Y_S-2 s_W^2+1\Big),\nn\\
\xi _{1,\gamma }&=-\frac{\left| y\right| ^2}{768 \pi ^2}\frac{s}{M^2}\left(2J_{S3}+2 Y_S+3\right), &
\xi _{1,Z} &=-\frac{\left| y\right| ^2}{768 \pi ^2 c_W s_W}\frac{s}{M^2}\Big(2 c_W^2 J_{S3}-2 s_W^2 Y_S-3 s_W^2+\frac{3}{2}-\frac{m_{\mu }^2}{s}\Big),\nn\\
\xi _{2,\gamma } &=+\frac{\left| y\right| ^2}{768 \pi ^2}\frac{m_{\mu }}{M^2}\left(4J_{S3}+4 Y_S+6\right),&
\xi _{2,Z}&=+\frac{\left| y\right| ^2}{768 \pi ^2 c_W s_W}\frac{m_{\mu }}{M^2}\Big(4 c_W^2 J_{S3}-4 s_W^2 Y_S-6 s_W^2+3\Big).
\end{align}
The $\xi_{1,\gamma}$ and $\xi_{2,\gamma}$ terms are correlated with respect to the electromagnetic current conservation~\cite{Stange:1993td,PhysRevD.79.015004} and appear as 
$$\xi _{1,\gamma}\gamma ^{\nu }\gamma _5 +  \xi _{2,\gamma}q^{\nu }\gamma _5 = \xi _{1,\gamma}(\gamma ^{\nu }  -2m_{\mu}/s~q^{\nu })\gamma _5,$$ 
which is  the so-called anapole moment term. The anapole moment $\xi_{1,\gamma}$ vanishes at $s=q^2=0$. We also see that the correction to $\gamma\mu^+\mu^-$ vertex in Eq.~\ref{eqn:mmlorentz} vanishes in the Thomson limit, i.e., $q^\mu\rightarrow0$ (and thus $s=q^2\rightarrow0$), as consistent with the electrical charge renormalization. 

\vspace*{3mm}
\subsubsection{\it The $S\bar{\mu}_RF$ coupling scenario}
\vspace*{3mm}
Now we consider the case that $F$ and $S$ couple to the SM right-handed singlet $\mu_R$ through the following Yukawa interaction,
\beq\label{yukr}
{\cal L}_{\rm Yuk}= ~y C_{ij} S^i {\bar{\mu}_R} F^j + h.c.~,
\eeq
where $C_{ij}=\langle I_{\mu_R} 0|I_SI_F;ij\rangle$, with $I_F=I_S$. The loop-induced anomalous couplings therefrom are similar to the $S\bar{\mu}_LF$ coupling scenario, because they come from the same sort of diagrams in Fig.~\ref{fig:mumu}.
Now  $ J_{F/S 3}\equiv\langle 00|\hat J_{S/F}^3|00\rangle= 0$, since $W^3\mu^+_R\mu^-_R$ vertex does not conserve the $SU(2)_L$ quantum number. We have $Y_S+Y_F=-1$ by the $U(1)_Y$ gauge symmetry. Choosing $Y_S$ as the independent quantum number, we present the full result of the anomalous couplings in terms of scalar functions in App.~\ref{app:yukawaright}.  In the approximation of large mass expansion, they become
\begin{align}
\alpha _{\gamma } &=+\frac{\left| y\right| ^2}{768 \pi ^2}\frac{s }{M^2}\left(2 Y_S+3\right),&
\alpha _Z&=+\frac{\left| y\right| ^2}{768 \pi ^2 c_W s_W}\frac{s }{M^2}\left(-2s_W^2  Y_S-3s_W^2 -\frac{m_{\mu }^2}{s}\right),\nn\\
\beta _{\gamma }&= +\frac{\left| y\right| ^2}{768 \pi ^2}\frac{m_{\mu }}{M^2}\left(4 Y_S+2\right),&
\beta _Z&=+\frac{\left| y\right| ^2}{768 \pi ^2 c_W s_W}\frac{m_{\mu }}{M^2}\left(-4 s_W^2 Y_S-2s_W^2 \right),\nn\\
\xi _{1,\gamma }&=+\frac{\left| y\right| ^2}{768 \pi ^2}\frac{s }{M^2}\left(2 Y_S+3\right), &
\xi _{1,Z} &=+\frac{\left| y\right| ^2}{768 \pi ^2 c_W s_W}\frac{s }{M^2}\left(-2s_W^2  Y_S-3s_W^2+\frac{m_{\mu }^2}{s}\right),\nn\\
\xi _{2,\gamma } &=-\frac{\left| y\right| ^2}{768 \pi ^2}\frac{m_{\mu }}{M^2}\left(4 Y_S+6\right),&
\xi _{2,Z}&=-\frac{\left| y\right| ^2}{768 \pi ^2 c_W s_W}\frac{m_{\mu }}{M^2} \left(-4s_W^2 Y_S-6s_W^2\right).
\end{align}
Note that the remarks following Eq.~\ref{ancp:mumuleft} also apply to the results above.

\subsection{Anomalous couplings induced by the purely gauge interaction}

\begin{figure}
    \centering
    \begin{subfigure}[b]{0.24\textwidth}
        \includegraphics[width=\textwidth]{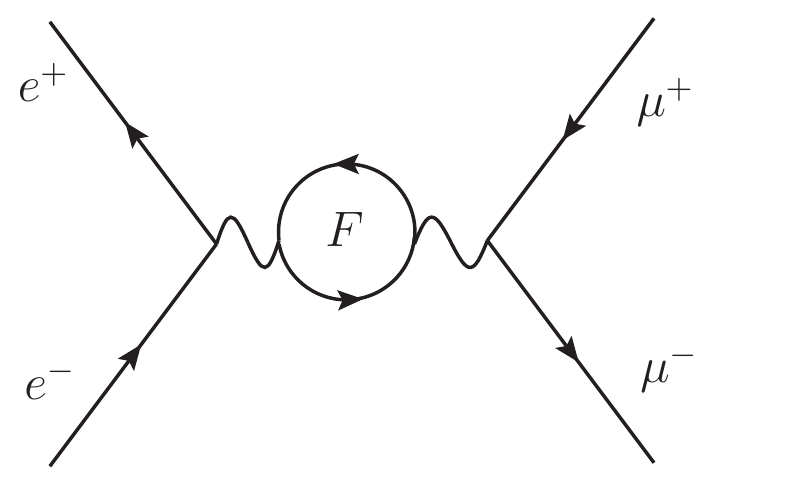}
        \subcaption{}\label{mumu: loop-p-s-F}
    \end{subfigure}
    \begin{subfigure}[b]{0.24\textwidth}
        \includegraphics[width=\textwidth]{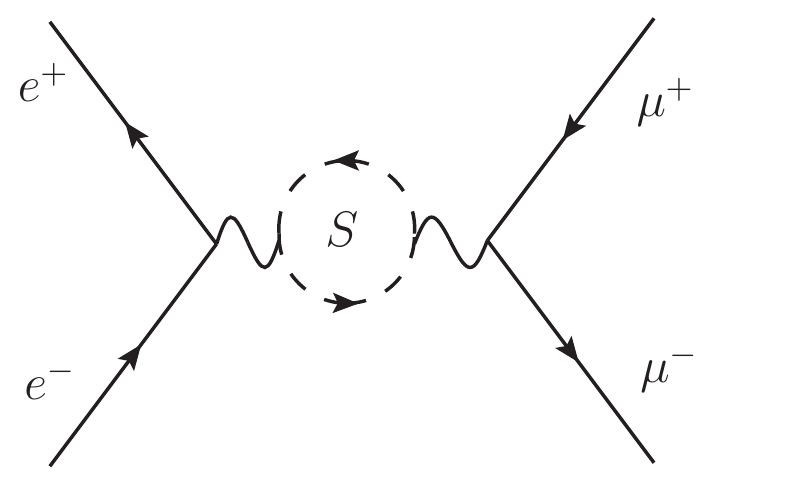}
        \subcaption{}\label{mumu: loop-p-s-S}
    \end{subfigure}
    \begin{subfigure}[b]{0.24\textwidth}
        \includegraphics[width=\textwidth]{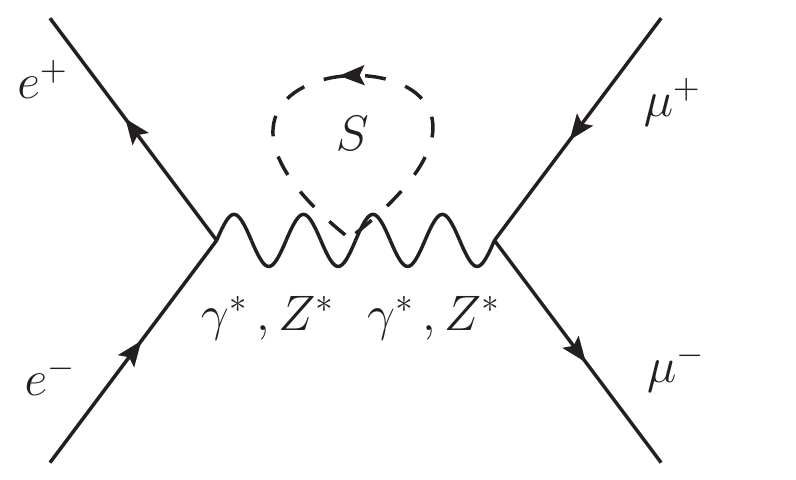}
        \subcaption{}\label{mumu: loop-p-s-S2}
    \end{subfigure}\\
    \begin{subfigure}[b]{0.24\textwidth}
        \includegraphics[width=\textwidth]{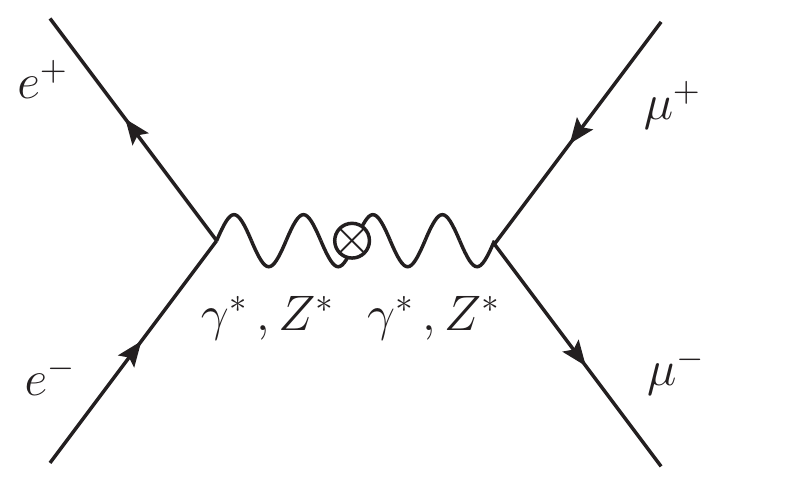}
        \subcaption{}\label{mumu: ct-p-s}
    \end{subfigure}
    \begin{subfigure}[b]{0.24\textwidth}
        \includegraphics[width=\textwidth]{ct-f-s.pdf}
        \subcaption{}\label{mumug: ct-f-s}
    \end{subfigure}
    \begin{subfigure}[b]{0.24\textwidth}
        \includegraphics[width=\textwidth]{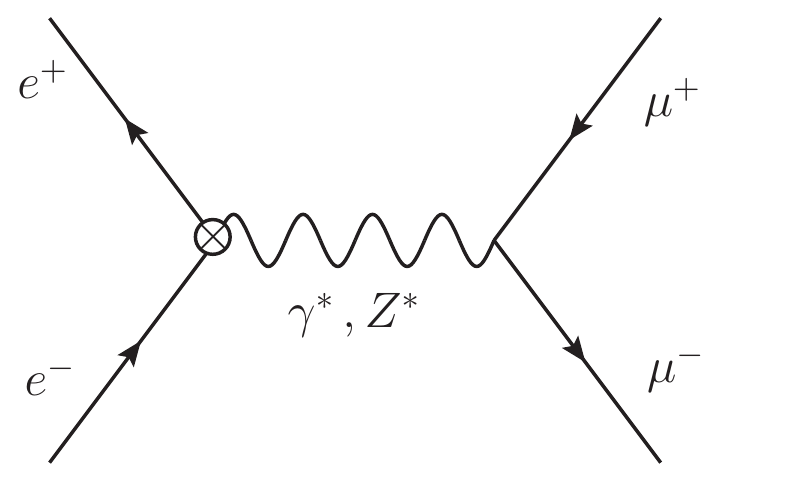}
        \subcaption{}\label{mumu: ct-i-s}
    \end{subfigure}
    \caption{Feynman diagrams of loop corrections to the $\eemm$ process from the gauge interaction.}
    \label{fig:mumugauge}
\end{figure}

The gauge interactions enter into the loop corrections of the $s-$channel propagators, as shown in Fig.~\ref{fig:mumugauge}. 
For convenience, we collect Fig.~\ref{mumu: loop-p-s-F} through Fig.~\ref{mumug: ct-f-s} and also parametrize the parts apart from the initial state matrix element as the $V\mu^+\mu^- $ anomalous couplings,
\beq\label{eqn:mmlorentzgauge}
-i e  \bar{u}\left(k_-\right) \left(\alpha _V\gamma ^{\nu }+\text{i$\beta $}_V  \sigma ^{\nu \rho }q_{\rho }+ \xi _{1,V}\gamma ^{\nu }\gamma _5 +  \xi _{2,V}q^{\nu }\gamma _5\right)v\left(k_+\right),
\eeq
where $V = \gamma\,, Z$.
As in Eq.~\ref{ffdecomp}, we decompose the couplings into the loop and counterterm parts,
\beq
\alpha_{V} =\alpha_{V,\bigcirc }+\delta \alpha_{V},\quad
\beta_{V} =\beta_{V,\bigcirc },\quad
\xi_{1,V} =\xi_{1,V,\bigcirc }+\delta \xi_{1,V},\quad 
\xi_{2,V} =\xi_{2,V,\bigcirc }+\delta \xi_{2,V},
\eeq
where the couplings with subscriptions $\bigcirc$ denote the contributions from the two-point loop corrections. The counterterm parts of the anomalous couplings are given as,
\begin{align}
\delta \alpha_{\gamma} &=\mathcal{C}^{\gamma}_{v}+\left( -\frac{g_{v} }{s-m_Z^2} \mathcal{C}_{A Z}+\frac{Q }{s} \mathcal{C}_{A A}\right),
&&\delta \alpha_{Z}~ =\mathcal{C}^{Z}_{v}+\left( -\frac{g_{v} }{s-m_Z^2} \mathcal{C}_{ZZ}+\frac{Q }{s} \mathcal{C}_{A Z}\right),\nn\\ 
\delta \xi_{1,\gamma} &=\mathcal{C}^{\gamma}_{a}+\left( -\frac{g_{a} }{s-m_Z^2} \mathcal{C}_{A Z}\right),
&&\delta \xi_{1,Z} =\mathcal{C}^{Z}_{a}+\left( -\frac{g_{a} }{s-m_Z^2} \mathcal{C}_{Z Z}\right),\nn\\  
\delta \xi_{2,\gamma} &=0~~+\left(~~\frac{g_a 2m_{\mu }}{s-m_Z^2}\mathcal{C}_{A Z}^\prime \right),
&&\delta \xi_{2,Z} =0~~+\left(~~\frac{g_a 2m_{\mu }}{s-m_Z^2}\mathcal{C}_{Z Z}^\prime \right),
\end{align}
where $g_v=(g_R+g_L)/2,~g_a=(g_R-g_L)/2$, and  the terms in the brackets come from the vector-vector counterterms depicted in Fig.~\ref{mumu: ct-p-s}, while the $\mathcal{C}_{v/a}^{\gamma/Z}$ terms are from the vertex counterterms shown in Fig.~\ref{mumug: ct-f-s}. Writing $\mathcal{C}^{\gamma/Z}_{v}=(\mathcal{C}^{\gamma/Z}_{R}+\mathcal{C}^{\gamma/Z}_{L})/2, ~\mathcal{C}^{\gamma/Z}_{a}=(\mathcal{C}^{\gamma/Z}_{R}-\mathcal{C}^{\gamma/Z}_{L})/2$, they are given as follows:
\begin{align}
\label{mmVcts}
&\mathcal{C}^{\gamma}_{L/R} =-Q \left(\frac{1}{2}\delta Z_{A A}+\delta Z_e\right)+ g_{L/R} \frac{1}{2}\delta Z_{{ZA}},\qquad \mathcal{C}^{Z}_{L/R} =g_{L/R} \left(\frac{\delta g_{L/R}}{g_{L/R}}+\frac{1}{2}\delta Z_{Z Z}\right) -Q\frac{1}{2} \delta Z_{{AZ}},\nn\\
&\mathcal{C}_{A A}= s \delta Z_{A A}, \qquad \mathcal{C}_{A Z}= \delta Z_{Z A} \left(s-m_Z^2\right)+s \delta Z_{A Z},\qquad \mathcal{C}_{Z Z}=\delta Z_{Z Z} \left(s-m_Z^2\right)-\delta m_Z^2,\nn\\
&\mathcal{C}_{A Z}^\prime= \frac{1}{2} \left({\delta Z}_{{AZ}}+{\delta Z}_{{ZA}}\right), \qquad\qquad\qquad\qquad\qquad \mathcal{C}_{Z Z}^\prime= {\delta Z}_{{ZZ}},
\end{align}
with 
\beq
\delta g_L=\frac{T^3}{-c_W s_W}\left[\frac{\delta s_W \left(s_W^2-c_W^2\right)}{c_W^2 s_W}+\delta Z_e\right]+\delta g_R,\quad \delta g_R=\frac{s_W}{c_W}Q\left[\frac{\delta s_W}{c_W^2 s_W}+\delta Z_e\right],
\eeq
where $\delta Z_{AA},~\delta Z_{ZA},~\delta Z_{AZ},~\delta Z_{ZZ}$,  $\delta Z_{e}$,  $\delta s_{W}/s_W$ and $\delta m_{Z}^2$ are the renormalization constants of wave function, electrical charge, weak mixing angle and the $Z$-boson mass, respectively.
Since $e^-$ and $\mu^-$ carry the same electroweak quantum numbers, the initial state counterterms in Fig.~\ref{mumu: ct-i-s} equal those in Fig.~\ref{mumug: ct-f-s}, and can be written as,
\beq\label{eqn:initalelectron}
-i e \gamma^\nu[\mathcal{C}^{\gamma/Z}_{L} P_L+ \mathcal{C}^{\gamma/Z}_{R}P_R],
\eeq
which are UV finite by themselves.

To renormalize the weak sector parameters, in the on-shell mass scheme we fix the mass and wave function renormalization constants by requiring that the renormalized parameters of the theory actually be equal to the physical parameters, i.e., the renormalized mass parameters be equal to the real parts of the poles of the corresponding propagators, and the residues of the propagators of the renormalized fields be equal to one. We further renormalize the electrical charge by equating it with the $ee\gamma $-coupling for on-shell external particles in the Thomson limit. In the on-shell scheme the weak mixing angle is a derived quantity.  We follow Sirlin's definition~\cite{Sirlin:1980nh} to define it as $c_{W}^{2} = m_{W}^{2}/m_{Z}^{2}$ using the renormalized gauge boson masses. To the  one-loop order we obtain
\beq
\frac{\delta c_{W}}{c_{W}} =\frac{1}{2}
\left(\frac{\delta m_{W}^{2}}{m_{W}^{2}}
-\frac{\delta m_{Z}^{2}}{m_{Z}^{2}}\right).
\eeq

Now we evaluate the loop diagrams in Fig.~\ref{mumu: loop-p-s-F} through~\ref{mumu: loop-p-s-S2}. Upon summing over the loop particle components, they are factorized into the corresponding generic self-energy integrals, multiplied by the group factors $C_F,~C_S$ and $D_F Y_F^2,~D_S Y_S^2$, where
\beq
\label{GF:CR}
C_S=\frac{1}{3}I_S(I_S+1)(2I_S+1),\qquad C_F=\frac{1}{3}I_F(I_F+1)(2I_F+1),
\eeq
are the Casimir invariants in representation $I_{S}$ and $I_{F}$ of the scalar $S$ and the fermion $F$, and $D_{S}=2I_{S}+1$ and $D_{F}=2I_{F}+1$ are their dimensions. As before, the generic self-energy integrals are reduced to one-loop scalar functions. We present the full result of the anomalous couplings in terms of scalar functions in App.~\ref{app:gaugeancoup}.  In the approximation of large mass expansion, they become
\begin{align}\label{ancp:gauge}
\alpha _{\gamma } &=-\frac{e^2}{3840 \pi ^2}\frac{s}{M^2}\frac{2}{c_W^2 s_W^2}\left[c_W^2 \left(8 C_F+C_S\right)+3 s_W^2 \left(8 D_F Y_F^2+D_S Y_S^2\right)\right],\nn\\
\xi _{1,\gamma }&=+\frac{e^2}{3840 \pi ^2}\frac{s}{M^2}\frac{2}{c_W^2 s_W^2}\left[c_W^2 \left(8 C_F+C_S\right)-s_W^2 \left(8 D_F Y_F^2+D_S Y_S^2\right)\right],\nn\\
\xi _{2,\gamma } &=-\frac{e^2}{3840 \pi ^2}\frac{m_{\mu }}{M^2}\frac{4}{c_W^2 s_W^2}\left[c_W^2 \left(8 C_F+C_S\right)-s_W^2 \left(8 D_F Y_F^2+D_S Y_S^2\right)\right], \nn\\
\alpha _Z &= -\frac{e^2}{3840 \pi ^2}\frac{s}{M^2}\frac{1}{c_W^3 s_W^3}\left[2 c_W^4 \left(8 C_F+C_S\right)-s_W^2 \left(8 D_F Y_F^2+D_S Y_S^2\right) \left(\frac{\left(2 c_W^2-3\right) m_Z^2}{s}+6 s_W^2\right)\right],\nn\\
\xi _{1,Z} &= +\frac{e^2}{3840 \pi ^2}\frac{s}{M^2}\frac{1}{c_W^3 s_W^3}\left[2 c_W^4 \left(8 C_F+C_S\right)+s_W^2 \left(8 D_F Y_F^2+D_S Y_S^2\right) \left(\frac{\left(2 c_W^2-1\right) m_Z^2}{s}+2 s_W^2\right)\right],\nn\\
\xi _{2,Z}&=-\frac{e^2}{3840 \pi ^2}\frac{m_{\mu }}{M^2}\frac{4}{c_W^3 s_W^3}\left[c_W^4 \left(8 C_F+C_S\right)+s_W^4 \left(8 D_F Y_F^2+D_S Y_S^2\right)\right]\frac{s-2 m_Z^2}{s-m_Z^2},\nn\\
\mathcal{C}^{Z}_{L}&= -\frac{e^2}{3840 \pi ^2}\frac{m_Z^2}{M^2}\frac{2}{c_W^3 s_W^3}\left[2 c_W^4 \left(8 C_F+C_S\right)+\left(2 c_W^2-1\right) s_W^2 \left(8 D_F Y_F^2+D_S Y_S^2\right)\right],\nn\\
\mathcal{C}^{Z}_{R}&= +\frac{e^2}{3840 \pi ^2}\frac{m_Z^2}{M^2}\frac{4}{c_W^3 s_W^3}\left[s_W^4 \left(8 D_F Y_F^2+D_S Y_S^2\right)\right],
\end{align}
and $\beta _{\gamma }= \beta _Z =0$. Notice that the purely counterterm corrections to $ee\gamma$ vertex $\mathcal{C}^{\gamma}_{L/R}=0$ exactly, since the electrical charge is renormalized to the $ee\gamma$ coupling strength at zero momentum transfer.

\section{Numerical results}

We choose our observable to be the deviation from the SM tree-level cross section 
$\Delta\sigma(\equiv \sigma-\sigma_0)$, 
where $\sigma_0$ stands for the SM tree-level cross sections~\footnote{The SM corrections to $\sigma_0$ have been calculated in Ref.~\cite{Passarino:1978jh,Bardin:1997xq,Hahn:2003ab}.} and $\sigma$ is the sum of the SM cross section and the new physics one-loop virtual corrections. Therefore, $\Delta \sigma$ is the cross section of the interference between the SM and the new physics virtual corrections. Note that this is both theoretically consistent, as the corrections to the cross sections are complete to this order in the perturbation series, and also numerically robust because the new physics one-loop amplitude squared is negligible compared to the interference contribution.  
Ignoring the electron mass, the correction $\Delta\sigma$ is given below in terms of the anomalous couplings,
\bea
&&\frac{{d\Delta\sigma }}{dt}=\frac{\pi \alpha_{\rm EM} ^2}{s^2}\sum\limits_{ij}\biggl[{\cal A}_{ij}^L\left(g_i^\gamma+g_Lg_i\frac{s}{s-m_Z^2}\right) \left( F^L_{\gamma, j}+F^L_{Z, j}\frac{s}{s-m_Z^2}\right)+\left(L\rightarrow R\right)\biggr].
\eea
The terms in the first round brackets come from the SM amplitudes, while those from the second round brackets come from the new physics loop corrections. $L$ and $R$ label the chirality of the initial state electrons (positrons). The index $i$, running through \{$v,~a$\}, labels the SM vector and axial-vector couplings of the final state $\mu^+\mu^-$ pair with $\gamma/Z$.  Note that $g_v^\gamma=1,~g_a^\gamma=0$. The index $j$, running through \{$\alpha,~\xi_1,~\beta$\}, labels the new physics loop-induced contributions, with
\begin{align}
F^L_{\gamma, j}&=2 \Re\left\{\alpha_{\gamma}+\mathcal{C}^{\gamma}_{L},~\xi _{1,\gamma},~m_\mu\beta _\gamma\right\},&&
F^L_{Z, j}=2 \Re\Big\{g_L\alpha_{Z}+\mathcal{C}^{Z}_{L}g_v,~g_L\xi _{1,Z}+\mathcal{C}^{Z}_{L}g_a,~g_Lm_\mu\beta _Z\Big\},\nn\\
F^R_{\gamma, j}&=2 \Re\left\{\alpha_{\gamma}+\mathcal{C}^{\gamma}_{R},~\xi _{1,\gamma},~m_\mu\beta _\gamma\right\},&&
F^R_{Z, j}=2 \Re\Big\{g_R\alpha_{Z}+\mathcal{C}^{Z}_{R}g_v,~g_R\xi _{1,Z}+\mathcal{C}^{Z}_{R}g_a,~g_Rm_\mu\beta _Z\Big\}.
\end{align}
Note that in the formula above, $\alpha_V,\beta_V,\xi_{1,V},\xi_{2,V}$ include both the Yukawa and the gauge corrections to the $V\mu^+\mu^-$ matrix elements; see Eqs.~\ref{eqn:mmlorentz} and \ref{eqn:mmlorentzgauge}.  $\mathcal{C}^{V}_{L/R}$ are from the counterterm corrections to the initial state $Ve^+e^-$ matrix elements; see Eqs.~\ref{eqn:initalelectron}.
The ${\cal A}$ functions are given by
\begin{align}
{\cal A}_{v,\alpha}^L&=+{\cal A}_{v,\alpha}^R=\frac{6 m_{\mu }^4-4 m_{\mu }^2 (t+u)+t^2+u^2}{s^2}, &&{\cal A}_{v,\xi_1 }^L=-{\cal A}_{v,\xi_1 }^R= \frac{u-t}{s},&&
{\cal A}_{v,\beta }^L=+{\cal A}_{v,\beta }^R= 2           ,\nn\\[1ex]
{\cal A}_{a,\alpha}^L&=-{\cal A}_{a,\alpha}^R=    \frac{u-t}{s}  ,&&
 {\cal A}_{a,\xi_1 }^L=+{\cal A}_{a,\xi_1 }^R=       \frac{-2 m_{\mu }^4+t^2+u^2}{s^2}  ,&&
{\cal A}_{a,\beta }^L=-{\cal A}_{a,\beta }^R=      \frac{2 (u-t)}{s} ,\nn
\end{align}
where $s=(p_-+p_+)^2$, $t=(p_--k_-)^2$, and $u=(p_- - k_+)^2$ are the usual Mandelstam variables.

Now we are ready to discuss our numerical results. The SM input parameters are chosen as follows~\cite{Mohr:2015ccw}: 
\beq
\begin{array}[b]{llll}\nn
 &G_\mu = 1.1663787\times10^{-5} \mbox{ GeV}^{-2}, \qquad& \alpha_{\rm EM}(0)=1/137.035 999 139 , \\[1ex]
 & m_Z = 91.1876 \mbox{ GeV}, \qquad& m_W = 80.385 \mbox{ GeV},  \qquad& m_\mu=105.658 3745 \mbox{ MeV},
\end{array}
\eeq
while the weak mixing angle is fixed by $c_W=m_W/m_Z$.
The loop corrections are calculated with the help of LoopTools package~\cite{Hahn:1998yk, vanOldenborgh:1990yc}. 
We choose the independent model parameters to be the Yukawa coupling strength $y$, the loop particle mass $M_F=M_S=M$, and the quantum numbers of the $S$ field $(I_S, Y_S)$.

\begin{figure}[b!]
        \includegraphics[scale=0.26]{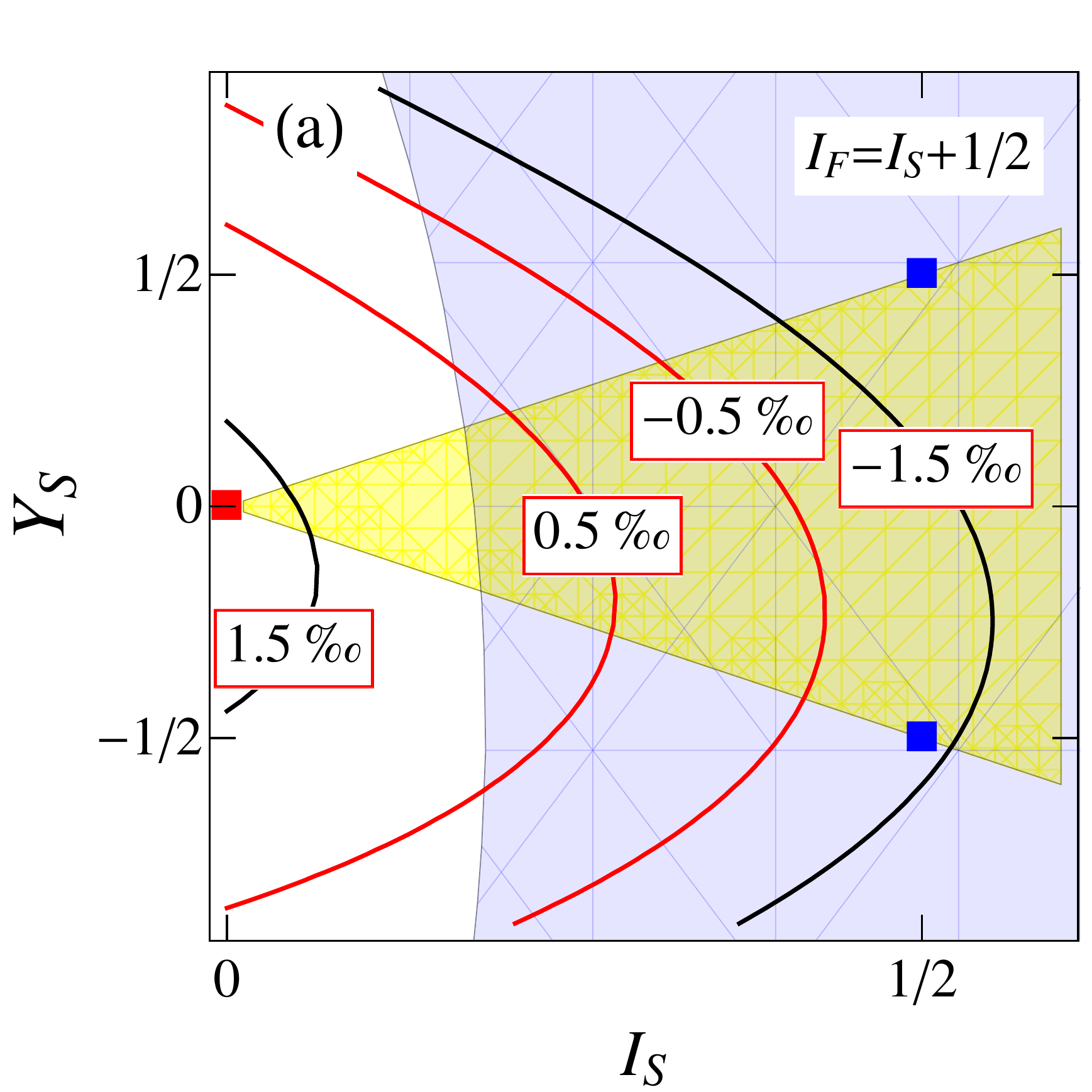}
        \includegraphics[scale=0.26]{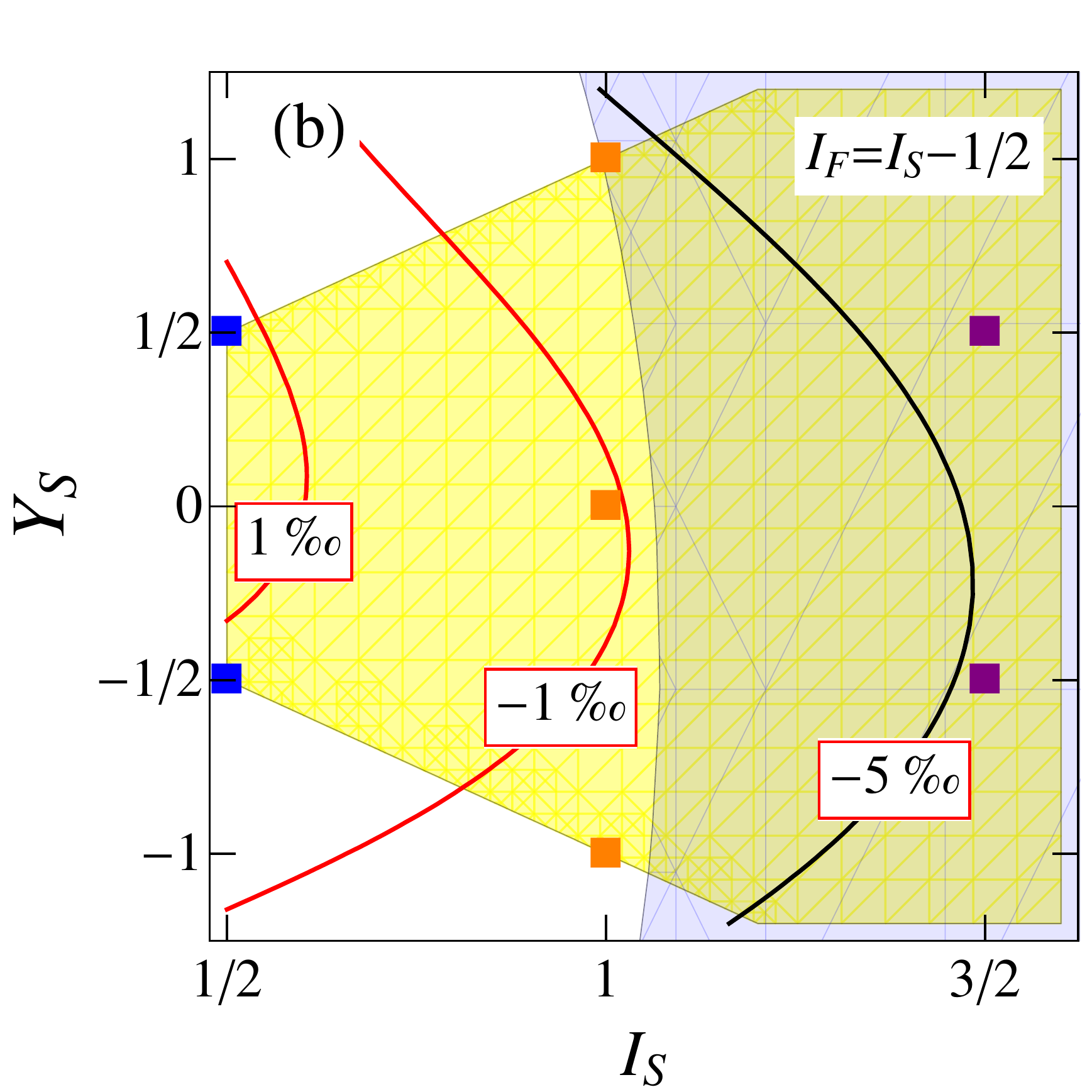}
        \includegraphics[scale=0.26]{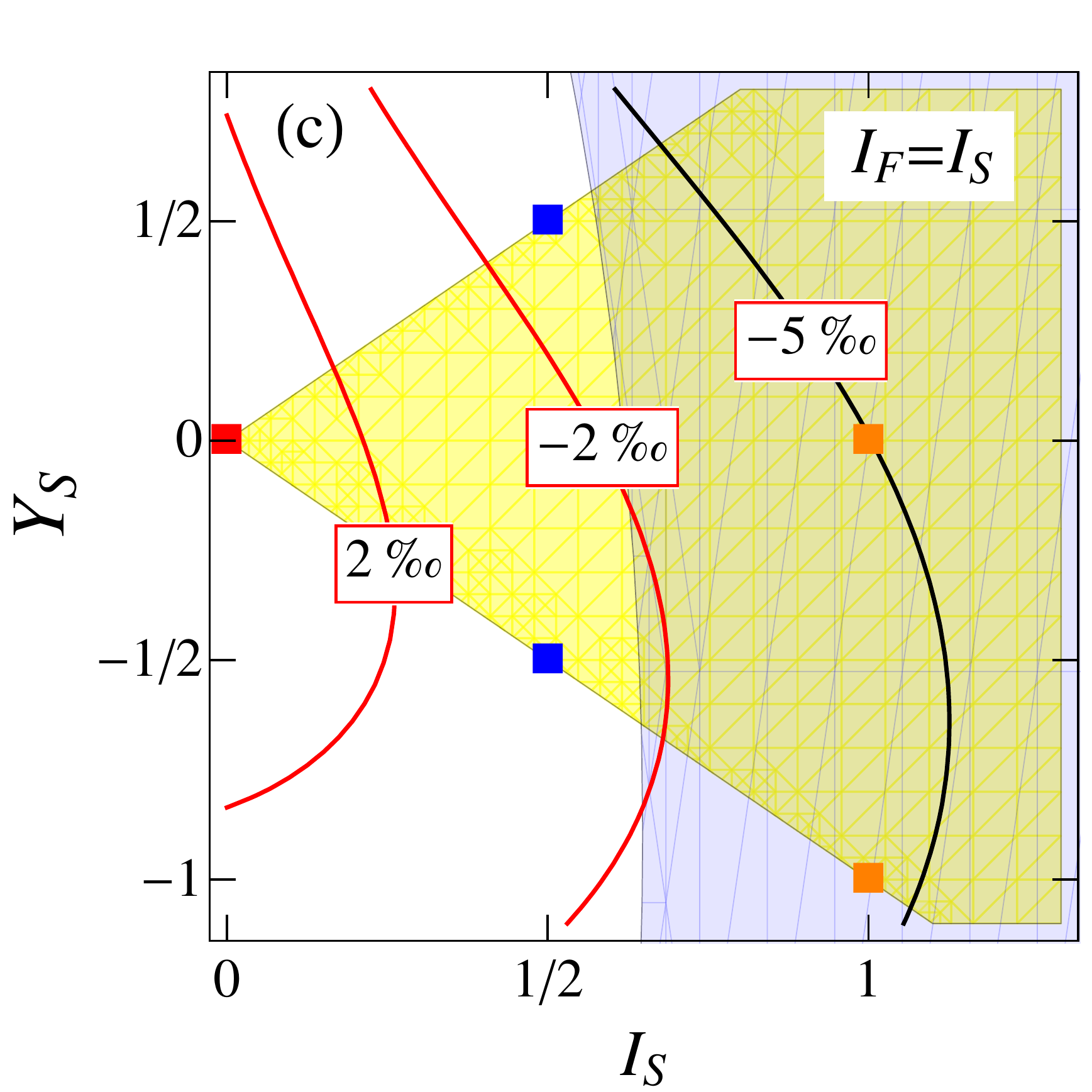}
\caption{Mono-jet data exclusion (blue region) of model quantum numbers of $F$ and $S$ fields with $M=150{\rm~ GeV}$: (a) the $S\bar{\mu}_LF$ coupling scenario with $I_F=I_S+1/2$, (b) the $S\bar{\mu}_LF$ coupling scenario with $I_F=I_S-1/2$, (c) the $S\bar{\mu}_RF$ coupling scenario. The yellow region encloses the model representations allowing for a scalar dark matter candidate. The contour lines help to estimate $\Delta\sigma/\sigma_0$ at CEPC for the various model representations with $y=1$.}
\label{fig:monojetmumu}
\end{figure}

First of all, we examine some possible experimental constraints on our model parameters. Consider the $S\bar{S}$ or $F\bar{F}$ pair productions at the LHC. In the degenerate-mass scenario, the SM decay products of $S$ or $F$, being very soft, can not be observed by detectors. To detect the $S\bar{S}$ or $F\bar{F}$ pairs, one has to make use of the jet ($j$) or the photon ($\gamma$) radiated out from the initial state partons, e.g. examining the process of $pp \to j + S\bar{S} (F\bar{F})$ or $pp \to \gamma+S\bar{S} (F\bar{F})$. That gives rise to a collider signature of one hard jet plus large $\met$ (named as mono-jet) or one hard photon plus large $\met$ (named as mono-photon), where the $\met$ originates from the $S\bar{S}$ and $F\bar{F}$ pairs. Therefore, the quantum numbers of $S$ and $F$ are constrained by the mono-jet or mono-photon data~\cite{Khachatryan:2014rra,Aad:2015zva,Aad:2014tda,Khachatryan:2014rwa}. We perform a simulation of the mono-jet and mono-photon productions using MadGraph5~\cite{Alwall:2011uj} with model files generated by FeynRules~\cite{Alloul:2013bka}, and find that the most stringent constraint comes from mono-jet experimental data, when $\met>400~{\rm GeV}$. The unfolded upper limits of new physics cross sections depend on the dark matter mass; for example, $\sigma(j+\met)\leq 0.76~{\rm pb}$ for a 150 GeV scalar dark matter particle. We choose $M=150$ GeV as a benchmark point, and apply the simulation results to constrain the quantum numbers $(I_S, Y_S)$. The 95\%~C.L. exclusion bounds are displayed in Fig.~\ref{fig:monojetmumu}; see the blue regions. The yellow regions enclose the model representations having an electrically neutral component as the dark matter candidate; see Eq.~\ref{modelregion}. The contour lines attached with relative correction values help to estimate the cross section corrections at the CEPC for the various model representations, with the Yukawa coupling strength chosen as $y=1$. Note that $I_S$ must be half integers, therefore, only those parameter points with box symbols represent realistic new physics models. The contour curves serve for the purpose of comparisons. We see that only a few lowest quantum numbers are allowed by the mono-jet data while the higher representations are excluded as they yield too much corrections. In the $S\bar{\mu}_LF$ coupling scenario, the highest allowed representation is a triplet scalar ($I_S=1$) and a doublet fermion ($I_F=1/2$), while in the $S\bar{\mu}_RF$ coupling scenario, the highest allowed representation is a double scalar and a doublet fermion ($I_S=I_F=1/2$).
Another important experimental constraint might come from the relic abundance measured by the Planck experiment~\cite{Ade:2015xua}, with $\Omega h^2=0.1186\pm 0.0020$, or equivalently, the thermally averaged annihilation cross section should be larger than $\langle\sigma v\rangle_{\rm Relic}\simeq 0.83~{\rm pb}$. In our interested parameter space region, the relic abundance constraint is easily satisfied~\cite{Burgess:2000yq,Cirelli:2005uq,LopezHonorez:2006gr,Hambye:2009pw,Agrawal:2014ufa}.

\begin{figure}
\centering
\includegraphics[scale=0.36]{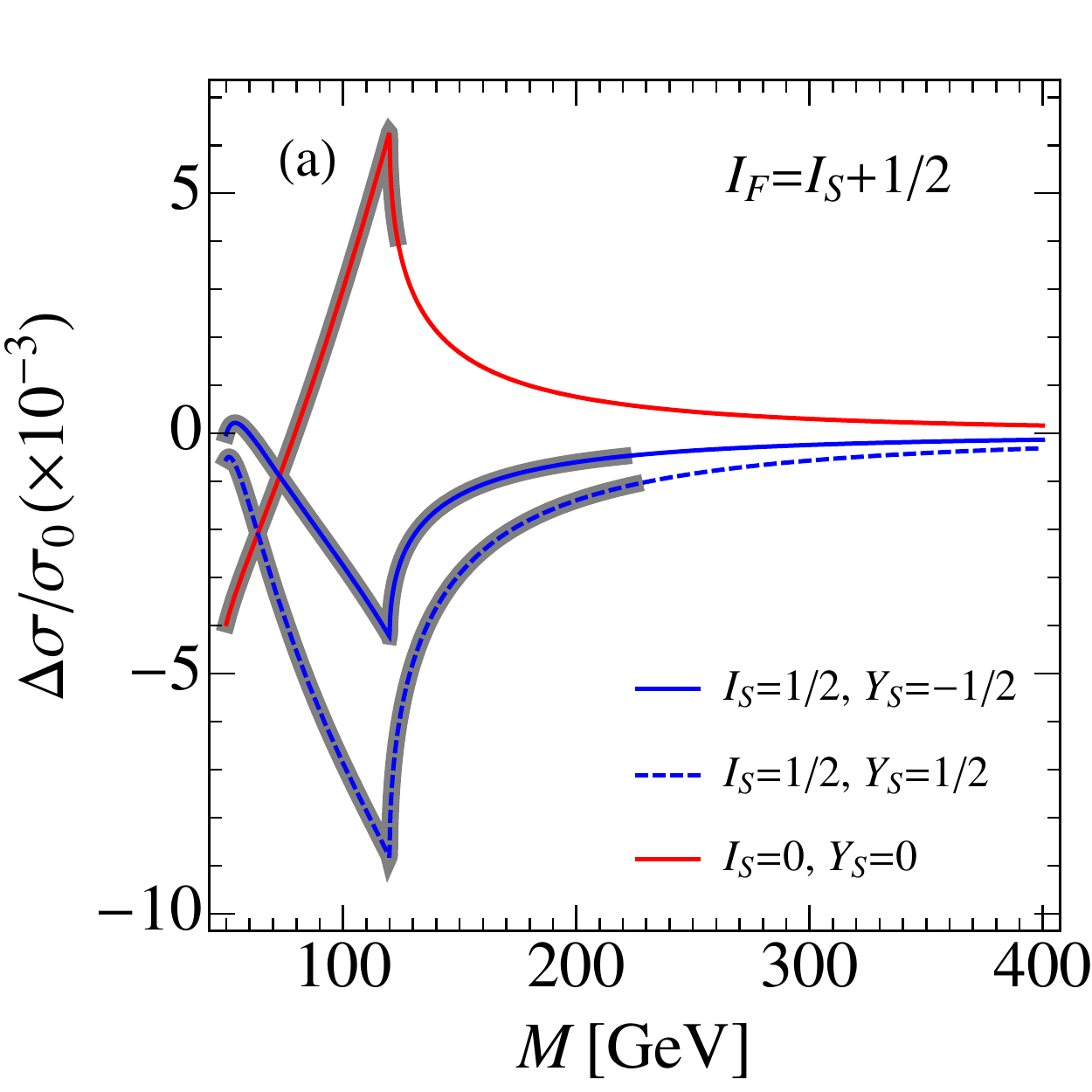}
\includegraphics[scale=0.36]{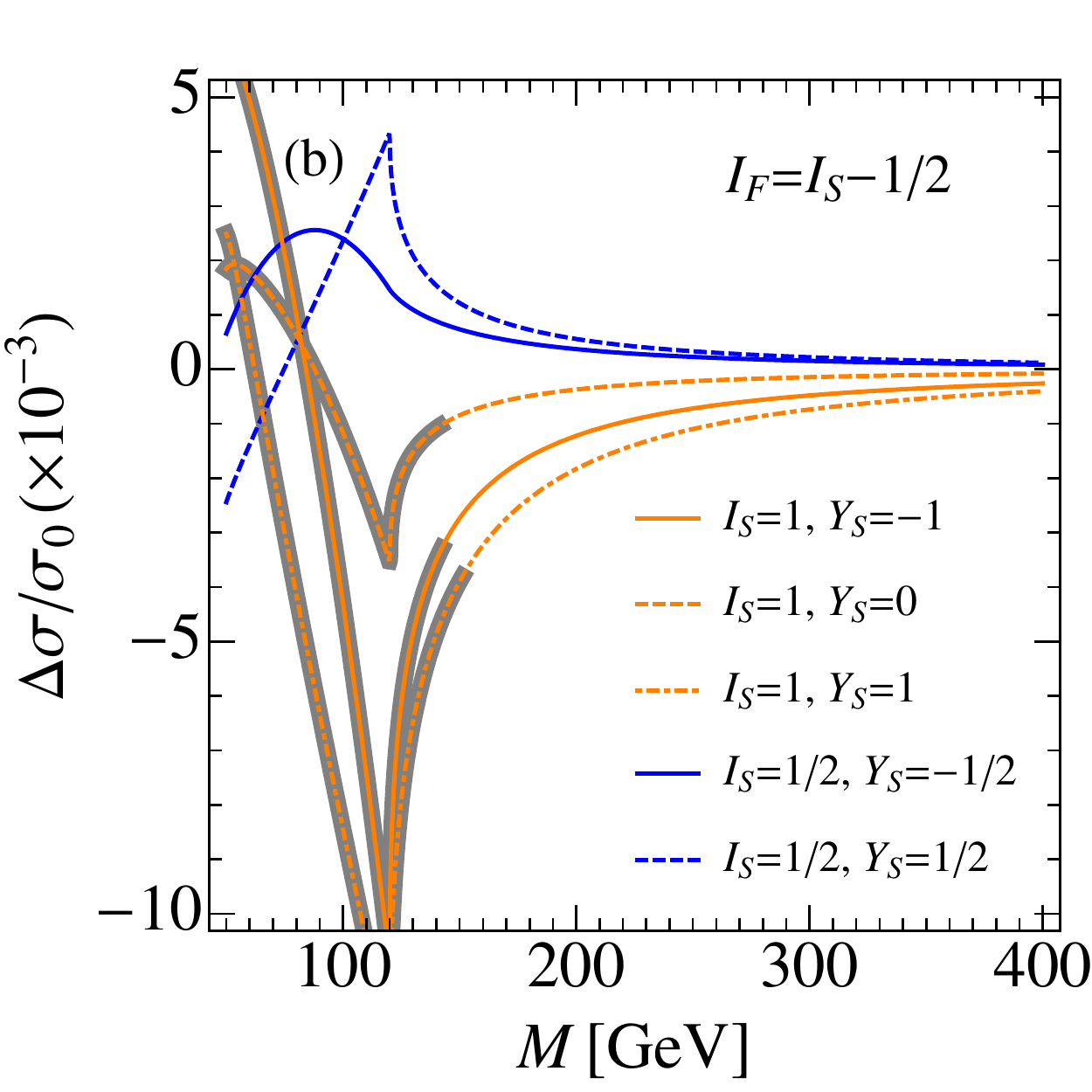}
\includegraphics[scale=0.36]{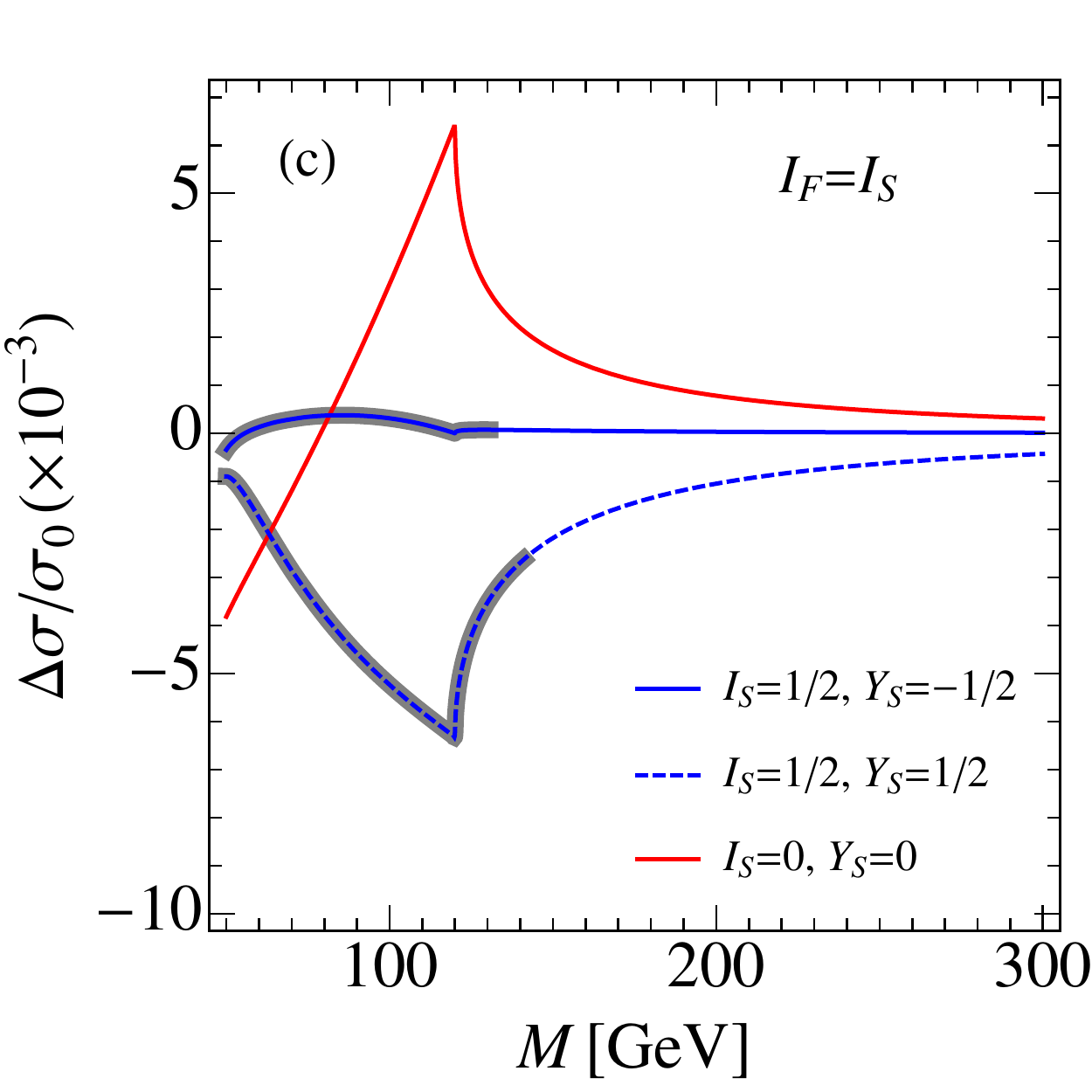}
\caption{Dependence of $\Delta\sigma/\sigma_0$ on $M$ at the CEPC with $y=1$: (a) the $S\bar{\mu}_LF$ coupling scenario with $I_F=I_S+1/2$, (b) the $S\bar{\mu}_LF$ coupling scenario with $I_F=I_S-1/2$, (c) the $S\bar{\mu}_RF$ coupling scenario. The grey bold part of of each curve is ruled out by the mono-jet data.}
\label{fig:mumulplots}
\end{figure}

In order to investigate the dependence of the loop particle mass, we choose $y=1$ as a benchmark point, and show in Fig.~\ref{fig:mumulplots} the corrections as a function of new physics particle mass $M$, for a few representations of $F$ and $S$. We have chosen $M>50{\rm~GeV}$ to eliminate the new physics correction to the weak gauge boson decay widths. The grey bold part of of each curve is ruled out by the mono-jet data. 
In the parameter space allowed by the mono-jet data, the largest correction comes from the $S\bar{\mu}_RF$ coupling scenario ($I_F=I_S$) at $M=120~{\rm GeV}$, increasing the SM prediction by about 6\textperthousand. The $S\bar{\mu}_LF$ coupling scenarios can increase the cross sections by 4\textperthousand. Assuming an accuracy of 2\textperthousand~at the CEPC, we notice a few points as follows: 
\begin{enumerate}
\item in the $S\bar{\mu}_LF$ coupling scenario with $I_F=I_S+1/2$ shown in Fig.~\ref{fig:mumulplots}(a), only the model with $I_S=0, Y_S=0$ (the red curve) can be testable at the CEPC; for example, a narrow mass window of about 30 GeV around $M=130~{\rm GeV}$ can yield a cross section deviation larger than 2\textperthousand;
\item in the $S\bar{\mu}_LF$ coupling scenario with $I_F=I_S-1/2$ displayed in Fig.~\ref{fig:mumulplots}(b), a relative positive corrections up to 3\textperthousand$\sim$4\textperthousand~can be yielded for $I_S=1/2, ~Y_S=\pm 1/2$ (the blue curves), whereas $I_S=1, Y_S=\pm1$ (the orange curves) give negative corrections up to about -4\textperthousand;
\item in the $S\bar{\mu}_RF$ coupling scenario depicted in Fig.~\ref{fig:mumulplots}(c), the model of $I_F=I_S=0, ~Y_S=0$ (the red curve) can be probed at the CEPC in the mass range of $90~{\rm GeV}\leq M \leq 140~{\rm GeV}$ and a maximal correction of 6\textperthousand~is achieved at 120 GeV.
\end{enumerate}
Therefore, we observe that the CEPC has a modest power to test certain model parameter space if 2\textperthousand~precision in the $\sigma(\eemm)$ measurement is achieved, but we point out that improving the accuracy to 1\textperthousand~shall enable us to probe a larger range of new physics mass.

\begin{figure}
\includegraphics[scale=0.35]{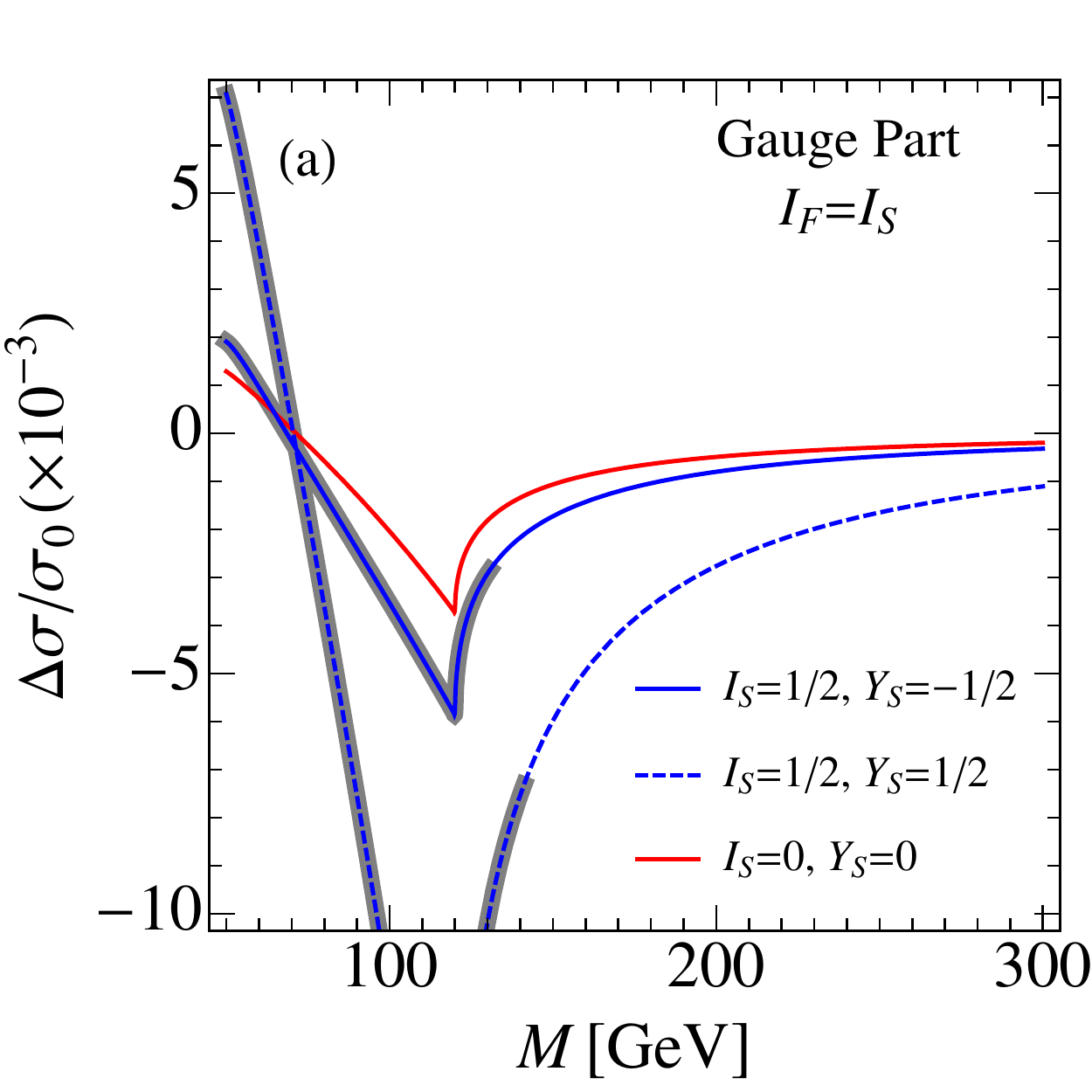}~~
\includegraphics[scale=0.35]{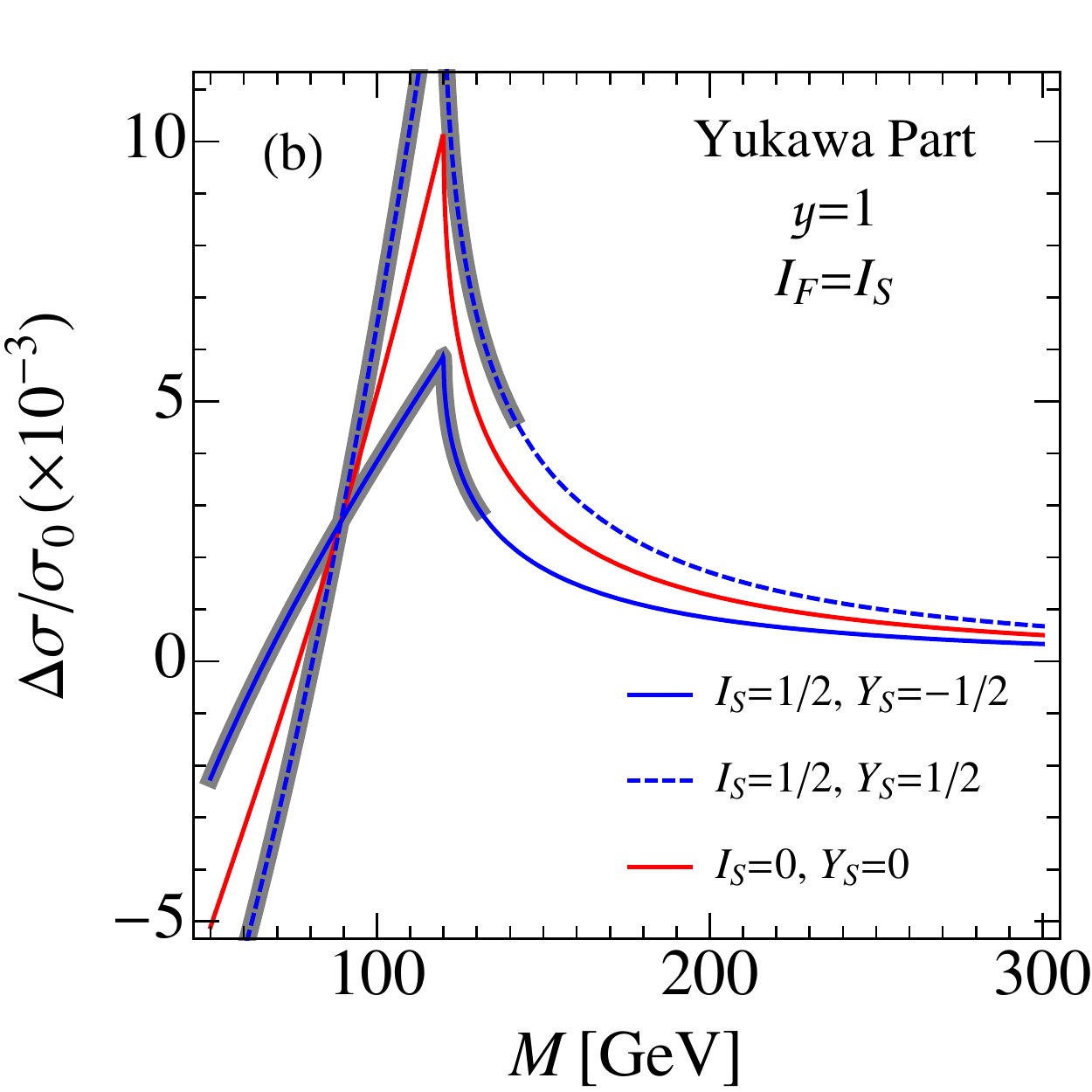}
\caption{The relative corrections from the purely gauge part (a) and the Yukawa part (b) in the $S\bar{\mu}_RF$ coupling scenario.}
\label{sepcon}
\end{figure}

The peak at the $M=\sqrt s/2= 120{\rm~GeV}$ for each curve in Fig.~\ref{fig:mumulplots} is due to the threshold effect from producing the intermediate on-shell $F\bar F$ pairs (c.f. Figs.~\ref{mumu: loop-f-s-F} and~\ref{mumu: loop-p-s-F}). The absence of such a peak for $I_S=1/2,~I_F=0,~Y_S=-1/2$ in the $S\bar{\mu}_LF$ coupling scenario, on the other hand, is because in this case only the scalar $S$ is coupled to $s-$channel gauge bosons (c.f. Figs.~\ref{mumu: loop-f-s-S} and~\ref{mumu: loop-p-s-S}), and the threshold effect is less pronounced, with the maximum shifted to a lower $M$ value. The sign of the cross section corrections exhibits a dependence on $I_S$ and $Y_S$. That is due to the interplay between the purely gauge correction and the purely Yukawa correction (to visualize the interplay, we display in Fig.~\ref{sepcon} both parts individually for the $S\bar{\mu}_RF$ coupling scenario $I_F=I_S$ as an example). Nevertheless, we also observe in addition to that a unanimous trend of negative corrections as the representations go higher, or as $I_S$ and $Y_S^2$ are larger. In such region of model parameter space, the gauge interaction corrections dominate over the Yukawa corrections, as can be readily seen from their power dependence in $I_S$ and $Y_S$ (c.f. Eqs.~\ref{GF:JS3},~\ref{ancp:mumuleft},~\ref{GF:CR} and~\ref{ancp:gauge}); therefore, the corrections approach to the (negative) purely gauge limit. 

 \begin{figure}
        \includegraphics[scale=0.35]{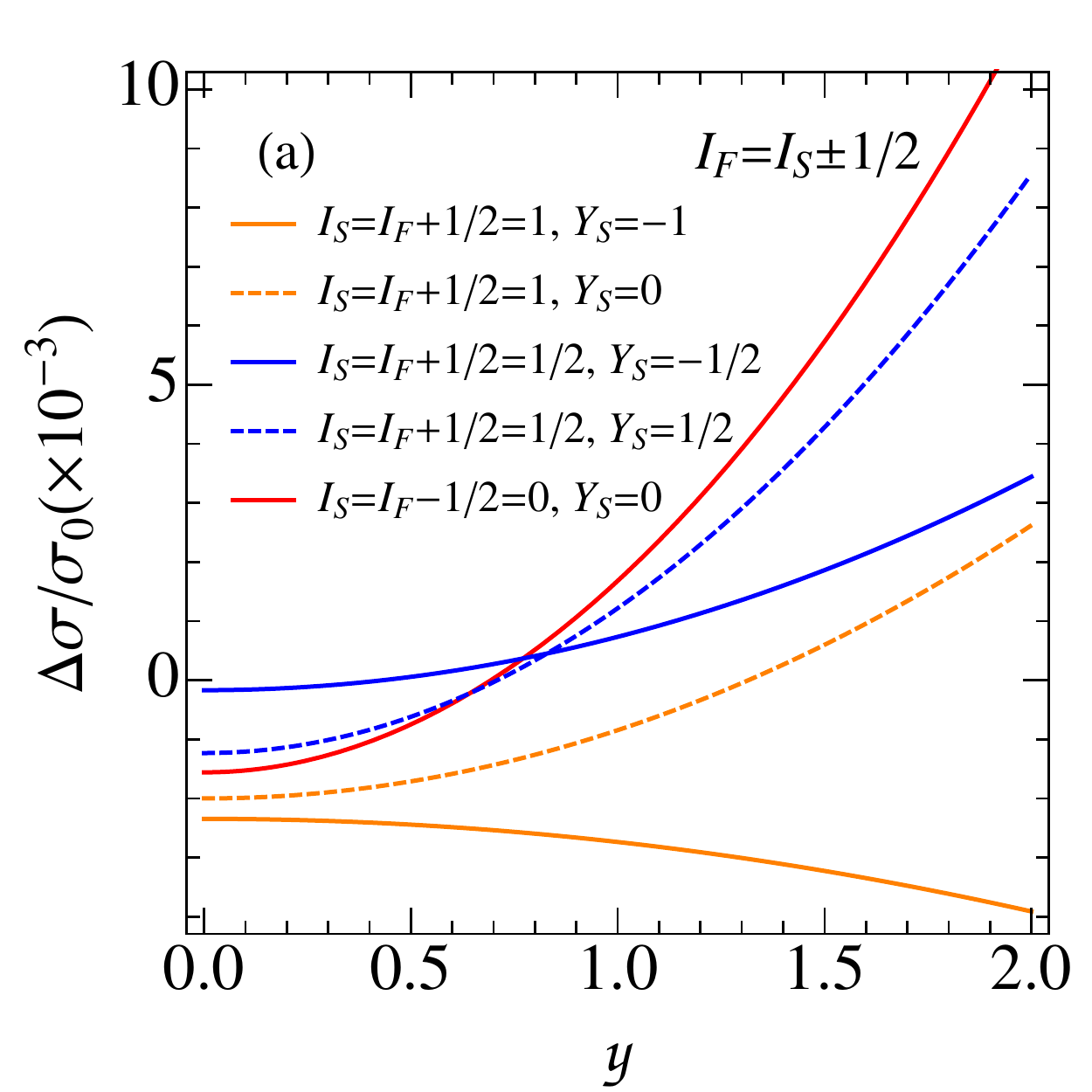}\qquad
        \includegraphics[scale=0.35]{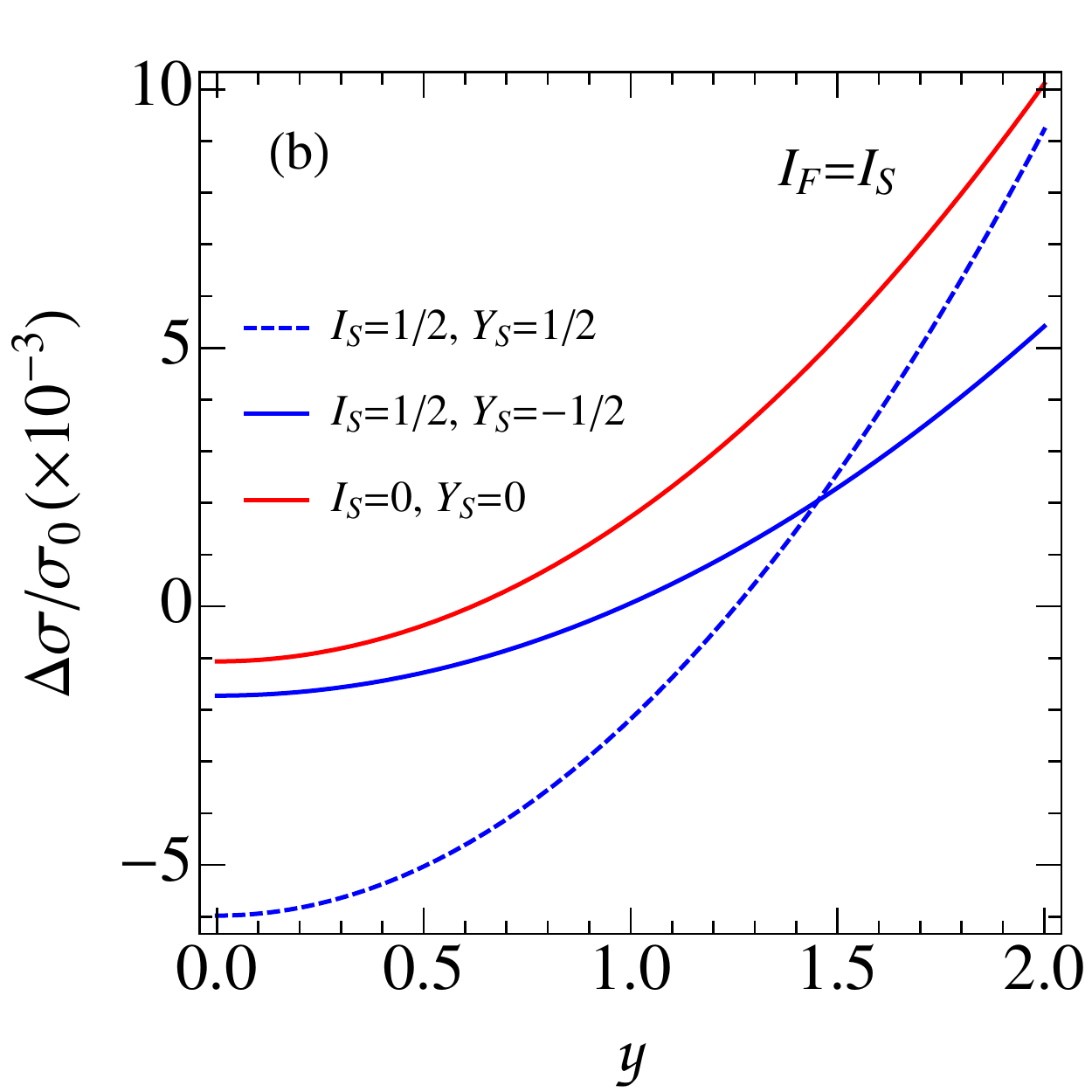}
\caption{The $y$ dependence of the relative corrections for the model quantum numbers allowed by mono-jet data in Fig.~\ref{fig:mumulplots} with $M=150$ GeV: (a) the $S\bar{\mu}_LF$ coupling scenario, (b) the $S\bar{\mu}_RF$ coupling  scenario.}
\label{ydependence}
\end{figure}

It is worth pointing out that, the corrections are sensitive to the Yukawa coupling strength $y$ through quadratic dependence $|y|^2$. We choose $M= 150$ GeV and plot in Fig.~\ref{ydependence} the relative corrections versus $y$ for those $I_S$ and $Y_S$ allowed by the mono-jet data; see Fig.~\ref{fig:mumulplots} for details. We observe that, except the model with $I_S=I_F+1/2=1,~Y_S=-1$ which has a negative contribution from the Yukawa part, the rest of the curves exhibit cancellation between the positive Yukawa part corrections and the negative gauge part corrections. For small values of $y$, the Yukawa part corrections become insignificant. The relative cross section corrections are dominated by the gauge corrections. The higher representations yield larger gauge corrections, which reaches -6\textperthousand~for $I_S=1/2,~Y_S=1/2$ in the $S\bar{\mu}_RF$ coupling scenario (c.f. the blue dashed curve in Fig.~\ref{ydependence}(b)). On the other hand, when $y$ is large, say $y\sim2$, the Yukawa part corrections dominate, and $\Delta\sigma/\sigma_0$ can reach above 1\% for both coupling scenarios (c.f. the red solid curves in Figs.~\ref{ydependence}(a) and \ref{ydependence}(b)).

\section{Conclusion}\label{sectionconclusion}

In this work, we addressed a ``nightmare" scenario in which the dark matter and its parent particles are nearly degenerate, so that the new physics signal would be difficult to probe at the LHC directly. However, the new physics particles affect the SM processes through quantum loop corrections, no matter whether they are degenerate or not, therefore, we proposed to explore their loop effects on $\sigma(\eemm)$ at the CEPC ($\sqrt{s}=240~{\rm GeV}$), with an expected accuracy of 2\textperthousand. 

In this work we payed our attentions to the case that one neutral component of the scalar particle $S$ serves as the dark matter candidate. A vector-like fermion multiplet $F$ has been introduced to facilitate the coupling of $S$ to the SM muon leptons through the Yukawa interaction. Various constraints from the mono-jet (photon) data and relic abundance are also discussed.

We have calculated the one-loop induced anomalous couplings of $\gamma\mu^+\mu^-$ and $Z\mu^+\mu^-$ for general $SU(2)_L\times U(1)_Y$ multiplets of $F$ and $S$. Our analytical results can be applied to many new physics models. Choosing Yukawa coupling strength $y=1$, the relative cross section corrections at the CEPC can reach above 2\textperthousand~for moderate new physics mass intervals and can be probed. For example, when the loop particle mass is around 120 GeV, the $S\bar{\mu}_RF$ coupling scenario with $I_S=Y_S=0$ can raise the SM cross section by +6\textperthousand,~and the $S\bar{\mu}_LF$ coupling scenario with $I_S=Y_S=1/2,~I_F=0$ can raise the SM cross section by +4\textperthousand. Furthermore, improving the accuracy to 1\textperthousand~would enable us to probe a larger range of new physics mass. We also discussed the relevance of the magnitude of $y$ and found that, for $M=150~{\rm GeV}$, when $y\ll1$, a negative correction of -6\textperthousand~can be reached in the $S\bar{\mu}_RF$ coupling scenario with $I_S=1/2,~Y_S=1/2$. When $y$ is large, say $y\sim2$, a positive correction of 1\% can be reached, for example, in the $S\bar{\mu}_LF$ coupling scenario with $I_S=Y_S=0, I_F=1/2$. Therefore, the nightmare scenario can be potentially examined at the CEPC.

\acknowledgments{
We thank Yandong Liu and Dong-Ming Zhang for useful discussions. The work is supported in part by the National Science Foundation of China under Grand No. 11275009.}

\appendix

\section{Feynman rules}		

\begin{figure}[h!]
    \centering
    \begin{subfigure}[b]{0.3\textwidth}
        \includegraphics[width=\textwidth]{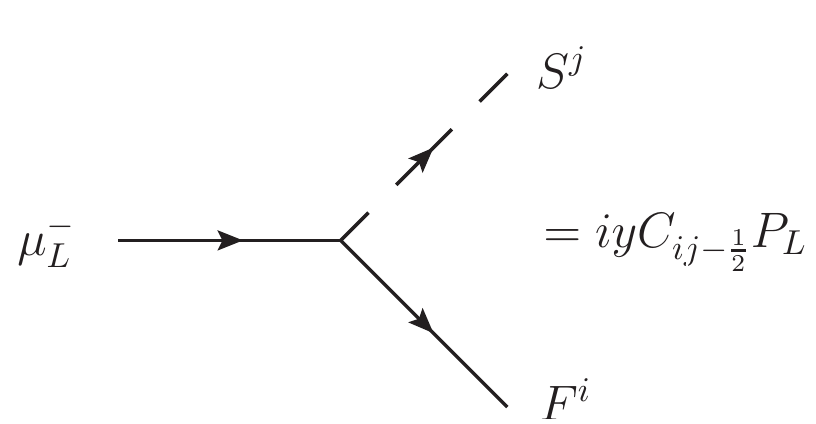}
        \subcaption{}
    \end{subfigure} \qquad
    \begin{subfigure}[b]{0.3\textwidth}
        \includegraphics[width=\textwidth]{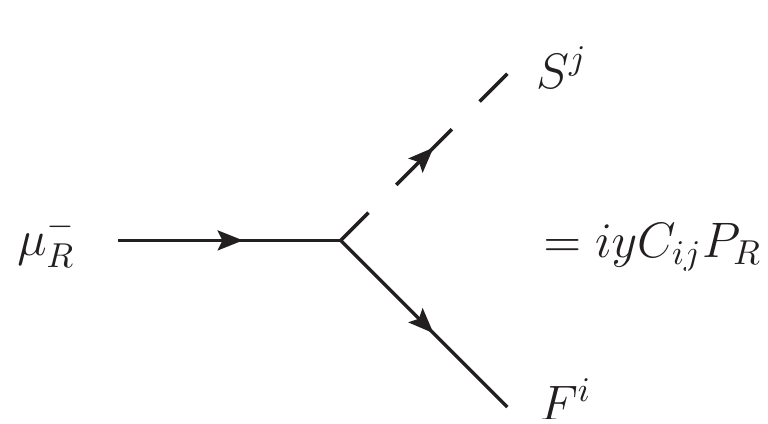}
        \subcaption{}
    \end{subfigure}
    \caption{Feynman rules of the $S\bar{\mu}_LF$ (a) and $S\bar{\mu}_RF$ (b) couplings in Eq.~\ref{yukl} and~\ref{yukr} respectively.}
    \label{fr-mumu-left-right}
\end{figure}

The Feynman rules for the Yukawa couplings in Eq.~\ref{yukl} and Eq.~\ref{yukr} are displayed in Fig.~\ref{fr-mumu-left-right} (a) and (b), respectively. The CG coefficients are given explicitly as follows,
\beq
\begin{split}
    C_{ij-\frac{1}{2}}&=
    \begin{cases}
      (-1)^{I_S-i} \sqrt{\dfrac{I_S+i+1}{2 I_S^2+3 I_S+1}}\delta_{i+j,-\frac{1}{2}},& \text{for  }I_F=I_S+\dfrac{1}{2},\\[3mm]
      (-1)^{I_S-i-1} \sqrt{\dfrac{I_S-i}{I_S (2 I_S+1)}}\delta_{i+j,-\frac{1}{2}}, & \text{for  }I_F=I_S-\dfrac{1}{2},
    \end{cases}
\end{split}
\eeq
and
\begin{align}
C_{ij}=(-1)^{I_S-i}{\frac{1}{\sqrt{2 I_S+1}}} \delta_{i+j,0}~~.
\end{align}

\section{Large mass expansion of scalar integrals}\label{app:lme}

With the Taylor series of Feynman integral denominator in large mass 
\begin{equation}
\frac{1}{(l+p)^2-M^2}=\frac{1}{l^2-M^2}\sum\limits_{n=0}^\infty \left(-\frac{2l\cdot p +p^2}{l^2-M^2}\right)^n, 
\end{equation}
we can expand the Passiano-Veltman scalar functions~\cite{Passarino:1978jh,tHooft:1978xw} $B_0$, $C_0$, with degenerate masses, as below, as appropriate for our anomalous couplings up to order $\mathcal{O}(M^{-2})$:
\begin{align}
&B_0(p^2;M^2,M^2)=B_0(0;M^2,M^2)+\frac{p^2}{6M^2}+\frac{p^4}{60M^4}+\mathcal{O}(M^{-6}),\nn\\
&C_0(p_1^2,p_2^2,p_3^2;M^2,M^2,M^2)=-\frac{1}{2M^2}-\frac{p_1^2+p_2^2+p_3^2}{24M^4}+\mathcal{O}(M^{-6}).
\end{align}

\section{Anomalous couplings in terms of scalar integrals}\label{app:ancoup}

\subsection{Anomalous couplings induced by the $S\bar{\mu}_LF$ interaction}\label{app:yukawaleft}
By introducing the following shorthand notations,
\begin{align}\label{scalarfunnotoations}
&B_{0}\left(p^{2}\right)\equiv B_{0}\left(p^{2};M^{2},M^{2}\right),\quad
B_{0}^{\prime}\left(q^2\right)\equiv\frac{\partial}{\partial p^{2}}\left.B_{0}\left(p^{2};M^{2},M^{2}\right)\right|_{p^{2}=q^2},\quad
C_{0}\equiv C_{0}\left(s,m_{\mu}^{2},m_{\mu}^{2};M^{2},M^{2},M^{2}\right),
\end{align}
the Yukawa part anomalous couplings $a=\alpha _{V },\beta _{V },\xi _{1,V },\xi _{2,V }$ induced by the $S\bar{\mu}_LF$ interaction, with $V=\gamma,Z$, are given by
\beq
a \equiv \frac{\left| y\right| ^2}{16\pi^{2}}\left[ a^{1}+a^{2}B_{0}\left(s\right)+a^{3}B_{0}\left(m_{\mu}^{2}\right)+a^{4}B_{0}^{\prime}\left(m_{\mu}^{2}\right) +a^{5}C_{0} \right].
\eeq
The nonzero $a^{i}=\alpha _{V }^i,\beta _{V }^i,\xi _{1,V }^i,\xi _{2,V }^i$, $i=1,\cdots,5$, are listed below,
\begin{flalign*}
&\alpha _{\gamma }^1=-\frac{s \left(2 Y_S+2 J_{S3}+1\right)}{4 \left(s-4 m_{\mu }^2\right)},&&
\alpha _{\gamma }^2=-\alpha _{\gamma }^3=\frac{16 m_{\mu }^4-2 s \left(6 J_{{S3}}+6 Y_S+7\right) m_{\mu }^2+s^2}{4 \left(s-4 m_{\mu }^2\right)^2},\\
&\alpha _{\gamma }^4=-\frac{m_{\mu }^2}{2},\\
\mathrlap{\alpha _{\gamma }^5=-\frac{s m_{\mu }^2 \left[s \left(J_{\text{S3}}+Y_S+1\right)-4 M^2 \left(2 J_{\text{S3}}+2 Y_S+1\right)\right]+s m_{\mu }^4 \left(2 J_{\text{S3}}+2 Y_S-3\right)+M^2 s^2 \left(2 J_{\text{S3}}+2 Y_S+1\right)+8 m_{\mu }^6}{2 \left(s-4 m_{\mu }^2\right){}^2},}\\
&\beta _{\gamma }^1=\frac{m_{\mu } \left(2 J_{{S3}}+2 Y_S+1\right)}{2 \left(s-4 m_{\mu }^2\right)},&&
\beta _{\gamma }^2=-\beta _{\gamma }^3=\frac{m_{\mu } \left(2 m_{\mu }^2+s\right) \left(2 J_{{S3}}+2 Y_S+1\right)}{2 \left(s-4 m_{\mu }^2\right)^2},\\
\mathrlap{\beta _{\gamma }^5=-\frac{m_{\mu } \left(2 J_{{S3}}+2 Y_S+1\right) \left[m_{\mu }^4+m_{\mu }^2 \left(4 M^2-s\right)-M^2 s\right]}{\left(s-4 m_{\mu }^2\right)^2},}\\
&\xi _{1,\gamma }^1=\frac{1}{4} \left(2 J_{{S3}}+2 Y_S+1\right),&&
\xi _{1,\gamma }^2=-\xi _{1,\gamma }^3=-\frac{\left(4 Y_S+4 J_{S3}+2\right) m_{\mu }^2+s}{4 \left(s-4 m_{\mu }^2\right)},\\
\mathrlap{\xi _{1,\gamma }^5=-\frac{m_{\mu }^2 \left[s \left(J_{{S3}}+Y_S+1\right)-m_{\mu }^2 \left(2 J_{{S3}}+2 Y_S+1\right)\right]-M^2
\left(s-4 m_{\mu }^2\right) \left(2 J_{{S3}}+2 Y_S+1\right)}{2\left(s-4 m_{\mu }^2\right)},}\\
&\mathrlap{\alpha _Z^1=-\frac{\left[4 c_W^2 J_{{S3}}-2 s_W^2 \left(2 Y_S+1\right) +1\right]s}{8c_Ws_W \left(s-4 m_{\mu }^2\right) },}\\
&\mathrlap{\alpha _Z^2=-\alpha _Z^3=\frac{2 m_{\mu }^2 s\left[2 s_W^2 \left(6 J_{\text{S3}}+6 Y_S+7\right)-12 J_{\text{S3}}-7\right]+\left(2 s_W^2-1\right) \left(16 m_{\mu }^4+s^2\right)}{8 c_W s_W \left(s-4 m_{\mu }^2\right)^2},}\\
&\alpha _Z^4=\frac{m_{\mu }^2 \left(4 s_W^2-1\right)}{8 c_W s_W},\\
\mathrlap{\alpha _Z^5=-\frac{2s \left[2 m_{\mu }^4+m_{\mu }^2 \left(s-8 M^2\right)+2 M^2 s\right] \left(c_W^2 J_{{S3}}-s_W^2 Y_S\right)+\left(c_W^2-s_W^2\right) \left[8 m_{\mu }^6+s m_{\mu }^2 \left(s-4 M^2\right)-3 s m_{\mu }^4+M^2 s^2\right]}{4 c_W s_W \left(s-4 m_{\mu }^2\right)^2},}\\
&\beta _Z^1=\frac{m_{\mu } \left[4 J_{{S3}}c_W^2-\left(4 Y_S+2\right) s_W^2+1\right]}{4c_W s_W\left(s- m_{\mu }^2\right) },&&
\beta _Z^2=-\beta _Z^3=-\frac{m_{\mu } \left(2 m_{\mu }^2+s\right) \left[\left(4 Y_S+2\right) s_W^2-4 J_{{S3}}c_W^2-1\right]}{4
c_W  s_W\left(s-4 m_{\mu }^2\right)^2},\\
\mathrlap{\beta _Z^5=\frac{m_{\mu } \left[m_{\mu }^4+\left(4 M^2-s\right) m_{\mu }^2-M^2 s\right] \left[\left(4 Y_S+2\right)s_W^2-4 J_{{S3}}c_W^2-1\right]}{2 c_W  s_W\left(s-4 m_{\mu }^2\right)^2},}\\
&\xi _{1,Z}^1=\frac{-2 \left(2 Y_S+1\right) s_W^2+4 J_{{S3}}c_W^2+1}{8 c_W s_W},&&
\xi _{1,Z}^2=-\xi _{1,Z}^3=\frac{2 \left[\left(4 Y_S+2\right) s_W^2-4 J_{{S3}}c_W^2-1\right] m_{\mu }^2+s \left(2 s_W^2-1\right)}{8c_Ws_W
\left( s- 4m_{\mu }^2\right) },\\
&\xi _{1,Z}^4=\frac{m_{\mu }^2}{8 c_W s_W},\\
\mathrlap{\xi _{1,Z}^5=-\frac{2(c_W^2 J_{{S3}}-s_W^2 Y_S) \left[-2 m_{\mu }^4+m_{\mu }^2 \left(8 M^2+s\right)-2 M^2 s\right]+\left(2 s_W^2-1\right) \left[m_{\mu }^4-m_{\mu }^2 \left(4 M^2+s\right)+M^2 s\right]}{4  c_W s_W \left(s-4 m_{\mu }^2\right)},}
\end{flalign*}
with $\xi _{2,V }^i=-\dfrac{2 m_{\mu }}{s}\xi _{1,V }^i$ for $V=\gamma\,,Z$ and $i\in\{1,2,3,4,5\}$, with the exception of $\xi _{2,Z}^4=0$.\\

\subsection{Anomalous couplings induced by the $S\bar{\mu}_RF$ interaction}\label{app:yukawaright}
Following the notations in~\ref{app:yukawaleft}, we list the nonzero scalar function coefficients of the anomalous couplings $\alpha _{V },\beta _{V },\xi _{1,V },\xi _{2,V }$ induced by the $S\bar{\mu}_LF$ interaction, with $V=\gamma,Z$, as follows,
\begin{flalign*}
&\alpha _{\gamma }^1=-\frac{s \left(2 Y_S+1\right)}{4 \left(s-4 m_{\mu }^2\right)},&
&\alpha _{\gamma }^2=-\alpha _{\gamma }^3=\frac{16 m_{\mu }^4-2 s \left(6 Y_S+7\right) m_{\mu }^2+s^2}{4 \left(s-4 m_{\mu }^2\right)^2},&&\alpha _{\gamma }^4=-\frac{m_{\mu }^2}{2},\\
\mathrlap{\alpha _{\gamma }^5=-\frac{8 m_{\mu }^6+s \left(2 Y_S-3\right) m_{\mu }^4+s \left[-4 M^2+s+\left(s-8 M^2\right) Y_S\right]
m_{\mu }^2+M^2 s^2 \left(2 Y_S+1\right)}{2 \left(s-4 m_{\mu }^2\right)^2},}\\
&\beta _{\gamma }^1=\frac{m_{\mu } \left(2 Y_S+1\right)}{2 \left(s-4 m_{\mu }^2\right)},&
&\beta _{\gamma }^2=-\beta _{\gamma }^3=\frac{m_{\mu } \left(2 m_{\mu }^2+s\right) \left(2 Y_S+1\right)}{2 \left(s-4 m_{\mu }^2\right)^2},\\
\mathrlap{\beta _{\gamma }^5=\frac{m_{\mu } \left[-m_{\mu }^4+\left(s-4 M^2\right) m_{\mu }^2+M^2 s\right] \left(2 Y_S+1\right)}{\left(s-4
m_{\mu }^2\right)^2},}\\
&\xi _{1,\gamma }^1=-\frac{1}{4} \left(2 Y_S+1\right),&
&\xi _{1,\gamma }^2=-\xi _{1,\gamma }^3=\frac{\left(4 Y_S+2\right) m_{\mu }^2+s}{4 \left(s-4 m_{\mu }^2\right)},\\
\mathrlap{\xi _{1,\gamma }^5=\frac{\left(s-4 m_{\mu }^2\right) \left(2 Y_S+1\right) M^2+m_{\mu }^2 \left[m_{\mu }^2 \left(2 Y_S+1\right)-s
\left(Y_S+1\right)\right]}{2\left(s-4 m_{\mu }^2\right)},}\\
&\alpha _Z^1=\frac{s s_W \left(2 Y_S+1\right)}{4 c_W \left(s-4 m_{\mu }^2\right)},&
&\alpha _Z^2=-\alpha _Z^3=-\frac{s_W \left[16 m_{\mu }^4-2 s \left(6 Y_S+7\right) m_{\mu }^2+s^2\right]}{4 c_W \left(s-4 m_{\mu }^2\right)^2},&&\alpha _Z^4=\frac{m_{\mu }^2 \left(4 s_W^2-1\right)}{8 c_W s_W},\\
\mathrlap{\alpha _Z^5=\frac{s_W \left\{8 m_{\mu }^6+s \left(2 Y_S-3\right) m_{\mu }^4+s \left[-4 M^2+s+\left(s-8 M^2\right) Y_S\right]
m_{\mu }^2+M^2 s^2 \left(2 Y_S+1\right)\right\}}{2 c_W \left(s-4 m_{\mu }^2\right)^2},}\\
&\beta _Z^1=-\frac{m_{\mu } s_W \left(2 Y_S+1\right)}{2 c_W \left(s-4 m_{\mu }^2\right)},&
&\beta _Z^2=-\beta _Z^3=-\frac{m_{\mu } \left(2 m_{\mu }^2+s\right) s_W \left(2 Y_S+1\right)}{2 c_W \left(s-4 m_{\mu }^2\right)^2},\\
\mathrlap{\beta _Z^5=\frac{m_{\mu } \left[m_{\mu }^4+\left(4 M^2-s\right) m_{\mu }^2-M^2 s\right] s_W \left(2 Y_S+1\right)}{c_W \left(s-4
m_{\mu }^2\right)^2},}\\
&\xi _{1,Z}^1=\frac{s_W \left(2 Y_S+1\right)}{4 c_W},&
&\xi _{1,Z}^2=-\xi _{1,Z}^3=-\frac{s_W \left[\left(4 Y_S+2\right) m_{\mu }^2+s\right]}{4c_W \left(s-4 m_{\mu }^2\right)},&&\xi _{1,Z}^4=\frac{m_{\mu }^2}{8 c_W s_W},\\
\mathrlap{\xi _{1,Z}^5=-\frac{s_W \left\{\left(4 m_{\mu }^2-s\right) \left(2 Y_S+1\right) M^2+m_{\mu }^2 \left[s \left(Y_S+1\right)-m_{\mu
}^2 \left(2 Y_S+1\right)\right]\right\}}{2c_W \left(s-4 m_{\mu }^2\right)},}
\end{flalign*}
with $\xi _{2,V }^i=-\dfrac{2 m_{\mu }}{s}\xi _{1,V }^i$ for $V=\gamma\,,Z$ and $i\in\{1,2,3,4,5\}$, with the exception of $\xi _{2,Z}^4=0$.\\

\subsection{Anomalous couplings induced by purely gauge interaction}\label{app:gaugeancoup}
Following the shorthand notations in Eq.~\ref{scalarfunnotoations} for the scalar functions, the gauge part anomalous couplings $a=\alpha _{V },\beta _{V },\xi _{1,V },\xi _{2,V }, \mathcal{C}_{L/R}^{V}$, with $V=\gamma,Z$, are given by
\beq
a\equiv\frac{e^2}{16\pi^{2}}\left[ a^{1}B_{0}\left(s\right)+a^{2}B_{0}\left(m_{W}^{2}\right)+a^{3}B_{0}\left(m_{Z}^{2}\right)+a^{4}B_{0}\left(0\right)+a^{5}B_{0}^{\prime}\left(m_{Z}^{2}\right)+a^{6}B_{0}^{\prime}\left(0\right) \right].
\eeq
The nonzero $a^{i}=\alpha _{V }^i,\beta _{V }^i,\xi _{1,V }^i,\xi _{2,V }^i, \mathcal{C}_{L/R}^{Vi}$, $i=1,\cdots,6$, are listed below,
\begin{align*}
\alpha_{{\gamma}}^{1}&=-\frac{\left(s-4 m_Z^2 s_W^2\right)\left[4 C_F \left(2 M^2+s\right)+C_S \left(s-4 M^2\right)\right]}{12 s s_W^2 \left(s-m_Z^2\right)}-\frac{\left(3 s-4 m_W^2\right)\left[4 D_F Y_F^2  \left(2 M^2+s\right)+D_S Y_S^2  \left(s-4 M^2\right)\right]}{12 s c_W^2 \left(s-m_Z^2\right)},\\
\alpha_{{\gamma}}^{3}&=\frac{s \left(4 s_W^2-1\right) \left\{c_W^2 \left[4 C_F \left(m_Z^2+2 M^2\right)+C_S \left(m_Z^2-4 M^2\right)\right]-s_W^2 \left[4 D_F Y_F^2  \left(m_Z^2+2 M^2\right)+D_S Y_S^2  \left(m_Z^2-4 M^2\right)\right]\right\}}{12 m_W^2 s_W^2 \left(m_Z^2-s\right)},\\
\alpha_{{\gamma}}^{4}&=\frac{C_F \left\{4 s_W^2 \left[m_Z^2 \left(2 M^2+s\right)+2 M^2 s\right]-2 M^2 s\right\}+C_S \left\{s_W^2 \left[m_Z^2 \left(s-4 M^2\right)-4 M^2 s\right]+M^2 s\right\}}{3 s m_Z^2 s_W^2}\\
&\quad+\frac{2 D_F Y_F^2  \left[2 m_W^2 \left(2 M^2+s\right)+M^2 s \left(1-4 s_W^2\right)\right]+D_S Y_S^2  \left[m_W^2 \left(s-4 M^2\right)+M^2 s \left(4 s_W^2-1\right)\right]}{3 s m_W^2},\\
\alpha_{{\gamma}}^{6}&=\frac{4}{3} M^2 \left(2 C_F-C_S+2 D_F Y_F^2 -D_S Y_S^2 \right),\\
\xi _{1,{\gamma}}^{1}&=\frac{c_W^2 \left[4 C_F \left(2 M^2+s\right)+C_S \left(s-4 M^2\right)\right]-s_W^2 \left[4 D_F Y_F^2  \left(2 M^2+s\right)+D_S Y_S^2  \left(s-4 M^2\right)\right]}{12 c_W^2 s_W^2 \left(s-m_Z^2\right)},\\
\xi _{1,{\gamma}}^{3}&=\frac{s \left\{c_W^2 \left[4 C_F \left(m_Z^2+2 M^2\right)+C_S \left(m_Z^2-4 M^2\right)\right]-s_W^2 \left[4 D_F Y_F^2  \left(m_Z^2+2 M^2\right)+D_S Y_S^2  \left(m_Z^2-4 M^2\right)\right]\right\}}{12 m_W^2 s_W^2 \left(m_Z^2-s\right)},\\
\xi _{1,{\gamma}}^{4}&=\frac{M^2 \left[c_W^2 \left(2 C_F-C_S\right)+s_W^2 \left(D_S Y_S^2 -2 D_F Y_F^2 \right)\right]}{3 m_W^2 s_W^2},\\
\alpha_{{Z}}^{1}&=-\frac{c_W \left(s-4 m_Z^2 s_W^2\right) \left[4 C_F \left(2 M^2+s\right)+C_S \left(s-4 M^2\right)\right]}{12 s s_W^3 \left(s-m_Z^2\right)}\\
&\quad+\frac{s_W \left(3 s-4 m_W^2\right) \left[4 D_F Y_F^2  \left(2 M^2+s\right)+D_S Y_S^2  \left(s-4 M^2\right)\right]}{12 s c_W^3 \left(s-m_Z^2\right)},\\
\alpha_{{Z}}^{2}&=\frac{\left(2 s_W^2+1\right) \left[4 C_F \left(m_W^2+2 M^2\right)+C_S \left(m_W^2-4 M^2\right)\right]}{24 c_W m_W^2 s_W^5},\\
\alpha_{{Z}}^{3}&=\frac{c_W^3 C_S \left[-m_Z^2 \left(16 M^2+s\right)+3 m_Z^4+8 M^2 s\right]}{24 m_W^2 s_W^3 \left(s-m_Z^2\right)}-\frac{c_W C_S  \left(m_Z^2-8 M^2+s\right)}{6  s_W \left(s-m_Z^2\right)}-\frac{c_W^3 C_S \left(m_Z^2-4 M^2\right)}{24 m_W^2 s_W^5}\\
&\quad+\frac{c_W^3 C_F \left[m_Z^2 \left(2 m_W^2+5 M^2-s\right)-3 M^2 s\right]}{3 m_W^2 s_W^3 \left(s-m_Z^2\right)}-\frac{c_W^5 C_F \left(m_Z^2+2 M^2\right)}{6 m_W^2 s_W^5}-\frac{2 c_W C_F  \left(4 M^2+s\right)}{3  s_W \left(s-m_Z^2\right)}\\
&\quad+\frac{D_S Y_S^2  \left\{c_W^2 \left(4 M^2-m_Z^2\right) \left(s-m_Z^2\right)+2 s_W^2 m_Z^2 \left[2 m_W^2+2 \left(8 M^2-s\right) s_W^2-10 M^2-s\right]+12 s_W^2  M^2 s\right\}}{24 c_W m_W^2 s_W \left(s-m_Z^2\right)}\\
&\quad-\frac{D_F Y_F^2  \left\{c_W^2 \left(m_Z^2+2 M^2\right) \left(s-m_Z^2\right)+2 s_W^2 m_Z^2 \left[-2 m_W^2+2 \left(4 M^2+s\right) s_W^2-5 M^2+s\right]+6 s_W^2 M^2 s\right\}}{6 c_W m_W^2 s_W \left(s-m_Z^2\right)},\\
\alpha_{{Z}}^{4}&=C_S \frac{s_W^2 \left\{m_W^2 \left[-4s s_W^2-32 M^2 c_W^2+s\right]+4 M^2 s \left(3-2 s_W^2\right)\right\}+8 M^2 s}{24 s c_W m_W^2 s_W^3}\\
&\quad+C_F \frac{s_W^2 \left\{m_W^2 \left[-4 s s_W^2+16 M^2c_W^2+s\right]+2 M^2 s \left(2 s_W^2-3\right)\right\}-4 M^2 s}{6 s c_W m_W^2 s_W^3}\\
&\quad+\frac{4 D_F Y_F^2  \left\{m_W^2 \left[s-4 \left(4 M^2+s\right) s_W^2\right]+2 M^2 s \left(2 s_W^2+1\right)\right\}}{24 s c_W m_W^2 s_W}\\
&\quad+\frac{D_S Y_S^2  \left\{m_W^2 \left[s-4 \left(s-8 M^2\right) s_W^2\right]-4 M^2 s \left(2 s_W^2+1\right)\right\}}{24 s c_W m_W^2 s_W},\\
\alpha_{{Z}}^{5}&=\frac{c_W \left(1-4 s_W^2\right) \left[4 C_F \left(m_Z^2+2 M^2\right)+C_S \left(m_Z^2-4 M^2\right)\right]}{24 s_W^3}\\
&\quad-\frac{s_W \left(4 s_W^2-1\right) \left[4 D_F Y_F^2  \left(m_Z^2+2 M^2\right)+D_S Y_S^2  \left(m_Z^2-4 M^2\right)\right]}{24 c_W^3},\\
\alpha_{{Z}}^{6}&=-\frac{M^2 \left(4 s_W^2-1\right) \left(2 C_F-C_S+2 D_F Y_F^2 -D_S Y_S^2 \right)}{6 c_W s_W},\\
\xi _{1,{Z}}^{1}&=\frac{c_W^4 \left[4 C_F \left(2 M^2+s\right)+C_S \left(s-4 M^2\right)\right]+s_W^4 \left[4 D_F Y_F^2  \left(2 M^2+s\right)+D_S Y_S^2  \left(s-4 M^2\right) \right]}{12 c_W^3 s_W^3 \left(s-m_Z^2\right)},\\
\xi _{1,{Z}}^{2}&=\frac{\left(2 s_W^2-1\right) \left[4 C_F \left(m_W^2+2 M^2\right)+C_S \left(m_W^2-4 M^2\right)\right]}{24 c_W m_W^2 s_W^5},\\
\xi _{1,{Z}}^{3}&=\left(c_W^4 C_S+s_W^4 D_S Y_S^2 \right)\frac{c_W^2 \left(4 M^2-m_Z^2\right) \left(s-m_Z^2\right)+2 s_W^2 \left[m_Z^2 \left(s-2 M^2\right)-2 M^2 s\right]}{24 c_W^3 m_Z^2 s_W^5 \left(m_Z^2-s\right)}\\
&\quad+\left(c_W^4 C_F+s_W^4 D_F Y_F^2 \right)\frac{2 s_W^2 \left[m_Z^2 \left(M^2+s\right)+M^2 s\right]-c_W^2 \left(m_Z^2+2 M^2\right) \left(s-m_Z^2\right)}{6 c_W^3 m_Z^2 s_W^5 \left(m_Z^2-s\right)},\\
\xi _{1,{Z}}^{4}&=\frac{C_S \left(20 M^2-c_W^2 m_Z^2\right)-4 C_F \left(c_W^2 m_Z^2+10 M^2\right)}{24 c_W^3 m_Z^2 s_W}+\frac{M^2 \left(s_W^4+1\right) \left(2 C_F-C_S\right)}{3 c_W^3 m_Z^2 s_W^3}\\
&\quad-\frac{4 D_F Y_F^2  \left[c_W^2 m_Z^2+2 M^2 \left(1-2 s_W^2\right)\right]+D_S Y_S^2  \left[c_W^2 m_Z^2+4 M^2 \left(2 s_W^2-1\right)\right]}{24 c_W^3 m_Z^2 s_W},\\
\xi _{1,{Z}}^{5}&=\frac{\left(4 M^2-m_Z^2\right) \left(c_W^4 C_S+s_W^4 D_S Y_S^2\right)}{24 c_W^3 s_W^3}-\frac{\left(m_Z^2+2 M^2\right) \left(c_W^4 C_F+s_W^4 D_F Y_F^2 \right)}{6 c_W^3 s_W^3},\\
\xi _{1,{Z}}^{6}&=\frac{M^2 \left(-2 C_F+C_S-2 D_F Y_F^2 +D_S Y_S^2 \right)}{6 c_W s_W},\\
\xi _{2,{Z}}^{1}&=-\frac{m_{\mu } \left(s-4 M^2\right) \left(c_W^4 C_S+D_S Y_S^2 s_W^4 \right)}{6 s c_W^3 s_W^3 \left(s-m_Z^2\right)}-\frac{2 m_{\mu } \left(2 M^2+s\right) \left(c_W^4 C_F+s_W^4 D_F Y_F^2 \right)}{3 s c_W^3 s_W^3 \left(s-m_Z^2\right)},\\
\xi _{2,{Z}}^{3}&=\frac{m_{\mu } \left[c_W^4 \left(4 C_F+C_S\right)+s_W^4 \left(4 D_F Y_F^2 +D_S Y_S^2 \right)\right]}{6 c_W^3 s_W^3 \left(s-m_Z^2\right)},\\
\xi _{2,{Z}}^{4}&=\frac{2 M^2 m_{\mu } \left[c_W^4 \left(2 C_F-C_S\right)+s_W^4 \left(2 D_F Y_F^2 -D_S Y_S^2 \right)\right]}{3 s c_W^3 s_W^3 \left(s-m_Z^2\right)},\\
\xi _{2,{Z}}^{5}&=\frac{m_{\mu } \left(m_Z^2-4 M^2\right) \left(c_W^4 C_S+D_S Y_S^2 s_W^4 \right)}{6 c_W^3 s_W^3 \left(s-m_Z^2\right)}+\frac{2 m_{\mu } \left(m_Z^2+2 M^2\right) \left(c_W^4 C_F+s_W^4 D_F Y_F^2 \right)}{3 c_W^3 s_W^3 \left(s-m_Z^2\right)},\\
\mathcal{C}_L^{Z2}&=\frac{4 C_F \left(m_W^2+2 M^2\right)+C_S \left(m_W^2-4 M^2\right)}{12 c_W m_W^2 s_W^5},\\
\mathcal{C}_L^{Z3}&=-c_W^3 \frac{4 C_F \left[m_Z^2 \left(2 s_W^4+s_W^2+1\right)+2 M^2 \left(4 s_W^4+1\right)\right]+C_S \left[m_Z^2 \left(2 s_W^4+s_W^2+1\right)-4 M^2 \left(4 s_W^4+1\right)\right]}{12 m_W^2 s_W^5}\\
&\quad+\left(2 s_W^2-1\right) \frac{4 D_F Y_F^2  \left[m_W^2+2 M^2 \left(1-2 s_W^2\right)\right]+D_S Y_S^2  \left[m_W^2+4 M^2 \left(2 s_W^2-1\right)\right]}{12 c_W m_W^2 s_W},\\
\mathcal{C}_L^{Z4}&=\frac{ \left(1-2 s_W^2\right) \left(4 C_F+C_S\right)}{12 c_W s_W}+\frac{M^2 \left(4 s_W^6-8 s_W^4+5 s_W^2-2\right) \left(2 C_F-C_S\right)}{3 c_W m_W^2 s_W^3}\\
&\quad-\left(2 s_W^2-1\right) \frac{4 D_F Y_F^2  \left[m_W^2+2 M^2 \left(1-2 s_W^2\right)\right]+D_S Y_S^2  \left[m_W^2+4 M^2 \left(2 s_W^2-1\right)\right]}{12 c_W m_W^2 s_W},\\
\mathcal{C}_L^{Z5}&=-\frac{\left(2 s_W^2-1\right) \left(4 M^2-m_Z^2\right) \left(c_W^4 C_S+s_W^4 D_S Y_S^2\right)}{12 c_W^3 s_W^3}+\frac{\left(2 s_W^2-1\right) \left(m_Z^2+2 M^2\right) \left(c_W^4 C_F+s_W^4 D_F Y_F^2 \right)}{3 c_W^3 s_W^3},\\
\mathcal{C}_L^{Z6}&=-\frac{M^2 \left(2 s_W^2-1\right) \left(2 C_F-C_S+2 D_F Y_F^2 -D_S Y_S^2 \right)}{3 c_W s_W},\\
\mathcal{C}_R^{Z2}&=\frac{4 C_F \left(m_W^2+2 M^2\right)+C_S \left(m_W^2-4 M^2\right)}{6 c_W m_W^2 s_W^3},\\
\mathcal{C}_R^{Z3}&=-c_W^3 \frac{4 C_F \left[m_Z^2 \left(s_W^2+1\right)+2 M^2 \left(2 s_W^2+1\right)\right]+C_S \left[m_Z^2 \left(s_W^2+1\right)-4 M^2 \left(2 s_W^2+1\right)\right]}{6 m_W^2 s_W^3}\\
&\quad+s_W \frac{4 D_F Y_F^2  \left[m_W^2+2 M^2 \left(1-2 s_W^2\right)\right]+D_S Y_S^2  \left[m_W^2+4 M^2 \left(2 s_W^2-1\right)\right]}{6 c_W m_W^2},\\
\mathcal{C}_R^{Z4}&=s_W \frac{C_S \left[-m_W^2+4 M^2 \left(3-2 s_W^2\right)\right]-4 C_F \left[m_W^2+2 M^2 \left(3-2 s_W^2\right)\right]}{6 c_W m_W^2}\\
&\quad+s_W \frac{4 D_F Y_F^2  \left[-m_W^2+2 M^2 \left(2 s_W^2-1\right)\right]+D_S Y_S^2  \left[-m_W^2+4 M^2 \left(1-2 s_W^2\right)\right]}{6 c_W m_W^2},\\
\mathcal{C}_R^{Z5}&=\frac{c_W^4 \left[4 C_F \left(m_Z^2+2 M^2\right)+C_S \left(m_Z^2-4 M^2\right)\right]+s_W^4 \left[4 D_F Y_F^2  \left(m_Z^2+2 M^2\right)+D_S Y_S^2  \left(m_Z^2-4 M^2\right)\right]}{6 c_W^3 s_W},\\
\mathcal{C}_R^{Z6}&=\frac{2 M^2 s_W \left(-2 C_F+C_S-2 D_F Y_F^2 +D_S Y_S^2 \right)}{3 c_W},
\end{align*}
with $\xi _{2,\gamma }^i=-\dfrac{2 m_{\mu }}{s}\xi _{1,\gamma }^i$ for $i\in\{1,2,3,4,5\}$.

\bibliographystyle{apsrev}
\bibliography{reference}

\begin{thebibliography}{47}
\expandafter\ifx\csname natexlab\endcsname\relax\def\natexlab#1{#1}\fi
\expandafter\ifx\csname bibnamefont\endcsname\relax
  \def\bibnamefont#1{#1}\fi
\expandafter\ifx\csname bibfnamefont\endcsname\relax
  \def\bibfnamefont#1{#1}\fi
\expandafter\ifx\csname citenamefont\endcsname\relax
  \def\citenamefont#1{#1}\fi
\expandafter\ifx\csname url\endcsname\relax
  \def\url#1{\texttt{#1}}\fi
\expandafter\ifx\csname urlprefix\endcsname\relax\def\urlprefix{URL }\fi
\providecommand{\bibinfo}[2]{#2}
\providecommand{\eprint}[2][]{\url{#2}}

\bibitem[{\citenamefont{Ahmed et~al.}(2010)}]{Ahmed:2009zw}
\bibinfo{author}{\bibfnamefont{Z.}~\bibnamefont{Ahmed}} \bibnamefont{et~al.}
  (\bibinfo{collaboration}{CDMS-II}), \bibinfo{journal}{Science}
  \textbf{\bibinfo{volume}{327}}, \bibinfo{pages}{1619} (\bibinfo{year}{2010}),
  \eprint{0912.3592}.

\bibitem[{\citenamefont{Bertone et~al.}(2005)\citenamefont{Bertone, Hooper, and
  Silk}}]{Bertone:2004pz}
\bibinfo{author}{\bibfnamefont{G.}~\bibnamefont{Bertone}},
  \bibinfo{author}{\bibfnamefont{D.}~\bibnamefont{Hooper}}, \bibnamefont{and}
  \bibinfo{author}{\bibfnamefont{J.}~\bibnamefont{Silk}},
  \bibinfo{journal}{Phys. Rept.} \textbf{\bibinfo{volume}{405}},
  \bibinfo{pages}{279} (\bibinfo{year}{2005}), \eprint{hep-ph/0404175}.

\bibitem[{\citenamefont{Akerib et~al.}(2014)}]{Akerib:2013tjd}
\bibinfo{author}{\bibfnamefont{D.~S.} \bibnamefont{Akerib}}
  \bibnamefont{et~al.} (\bibinfo{collaboration}{LUX}), \bibinfo{journal}{Phys.
  Rev. Lett.} \textbf{\bibinfo{volume}{112}}, \bibinfo{pages}{091303}
  (\bibinfo{year}{2014}), \eprint{1310.8214}.

\bibitem[{\citenamefont{Langacker}(2009)}]{Langacker:2008yv}
\bibinfo{author}{\bibfnamefont{P.}~\bibnamefont{Langacker}},
  \bibinfo{journal}{Rev. Mod. Phys.} \textbf{\bibinfo{volume}{81}},
  \bibinfo{pages}{1199} (\bibinfo{year}{2009}), \eprint{0801.1345}.

\bibitem[{\citenamefont{Olive et~al.}(2014)}]{Agashe:2014kda}
\bibinfo{author}{\bibfnamefont{K.~A.} \bibnamefont{Olive}} \bibnamefont{et~al.}
  (\bibinfo{collaboration}{Particle Data Group}), \bibinfo{journal}{Chin.
  Phys.} \textbf{\bibinfo{volume}{C38}}, \bibinfo{pages}{090001}
  (\bibinfo{year}{2014}).

\bibitem[{\citenamefont{Riemann}(2001)}]{Riemann:2001bb}
\bibinfo{author}{\bibfnamefont{S.}~\bibnamefont{Riemann}}, pp.
  \bibinfo{pages}{1451--1468} (\bibinfo{year}{2001}).

\bibitem[{\citenamefont{Baer et~al.}(2013)\citenamefont{Baer, Barklow, Fujii,
  Gao, Hoang, Kanemura, List, Logan, Nomerotski, Perelstein
  et~al.}}]{Baer:2013cma}
\bibinfo{author}{\bibfnamefont{H.}~\bibnamefont{Baer}},
  \bibinfo{author}{\bibfnamefont{T.}~\bibnamefont{Barklow}},
  \bibinfo{author}{\bibfnamefont{K.}~\bibnamefont{Fujii}},
  \bibinfo{author}{\bibfnamefont{Y.}~\bibnamefont{Gao}},
  \bibinfo{author}{\bibfnamefont{A.}~\bibnamefont{Hoang}},
  \bibinfo{author}{\bibfnamefont{S.}~\bibnamefont{Kanemura}},
  \bibinfo{author}{\bibfnamefont{J.}~\bibnamefont{List}},
  \bibinfo{author}{\bibfnamefont{H.~E.} \bibnamefont{Logan}},
  \bibinfo{author}{\bibfnamefont{A.}~\bibnamefont{Nomerotski}},
  \bibinfo{author}{\bibfnamefont{M.}~\bibnamefont{Perelstein}},
  \bibnamefont{et~al.} (\bibinfo{year}{2013}), \eprint{1306.6352}.

\bibitem[{\citenamefont{Group}(2015)}]{CEPC-SPPCStudyGroup:2015csa}
\bibinfo{author}{\bibfnamefont{C.-S.~S.} \bibnamefont{Group}}
  (\bibinfo{year}{2015}).

\bibitem[{\citenamefont{Jungman et~al.}(1996)\citenamefont{Jungman,
  Kamionkowski, and Griest}}]{Jungman:1995df}
\bibinfo{author}{\bibfnamefont{G.}~\bibnamefont{Jungman}},
  \bibinfo{author}{\bibfnamefont{M.}~\bibnamefont{Kamionkowski}},
  \bibnamefont{and} \bibinfo{author}{\bibfnamefont{K.}~\bibnamefont{Griest}},
  \bibinfo{journal}{Phys. Rept.} \textbf{\bibinfo{volume}{267}},
  \bibinfo{pages}{195} (\bibinfo{year}{1996}), \eprint{hep-ph/9506380}.

\bibitem[{\citenamefont{Fox and Poppitz}(2009)}]{Fox:2008kb}
\bibinfo{author}{\bibfnamefont{P.~J.} \bibnamefont{Fox}} \bibnamefont{and}
  \bibinfo{author}{\bibfnamefont{E.}~\bibnamefont{Poppitz}},
  \bibinfo{journal}{Phys. Rev.} \textbf{\bibinfo{volume}{D79}},
  \bibinfo{pages}{083528} (\bibinfo{year}{2009}), \eprint{0811.0399}.

\bibitem[{\citenamefont{Cao et~al.}(2009{\natexlab{a}})\citenamefont{Cao, Ma,
  and Shaughnessy}}]{Cao:2009yy}
\bibinfo{author}{\bibfnamefont{Q.-H.} \bibnamefont{Cao}},
  \bibinfo{author}{\bibfnamefont{E.}~\bibnamefont{Ma}}, \bibnamefont{and}
  \bibinfo{author}{\bibfnamefont{G.}~\bibnamefont{Shaughnessy}},
  \bibinfo{journal}{Phys. Lett.} \textbf{\bibinfo{volume}{B673}},
  \bibinfo{pages}{152} (\bibinfo{year}{2009}{\natexlab{a}}),
  \eprint{0901.1334}.

\bibitem[{\citenamefont{Agrawal et~al.}(2014)\citenamefont{Agrawal, Chacko, and
  Verhaaren}}]{Agrawal:2014ufa}
\bibinfo{author}{\bibfnamefont{P.}~\bibnamefont{Agrawal}},
  \bibinfo{author}{\bibfnamefont{Z.}~\bibnamefont{Chacko}}, \bibnamefont{and}
  \bibinfo{author}{\bibfnamefont{C.~B.} \bibnamefont{Verhaaren}},
  \bibinfo{journal}{JHEP} \textbf{\bibinfo{volume}{08}}, \bibinfo{pages}{147}
  (\bibinfo{year}{2014}), \eprint{1402.7369}.

\bibitem[{\citenamefont{Akerib et~al.}(2015)}]{Akerib:2015rjg}
\bibinfo{author}{\bibfnamefont{D.~S.} \bibnamefont{Akerib}}
  \bibnamefont{et~al.} (\bibinfo{collaboration}{LUX}) (\bibinfo{year}{2015}),
  \eprint{1512.03506}.

\bibitem[{\citenamefont{Chen and Takahashi}(2009)}]{Chen:2008dh}
\bibinfo{author}{\bibfnamefont{C.-R.} \bibnamefont{Chen}} \bibnamefont{and}
  \bibinfo{author}{\bibfnamefont{F.}~\bibnamefont{Takahashi}},
  \bibinfo{journal}{JCAP} \textbf{\bibinfo{volume}{0902}}, \bibinfo{pages}{004}
  (\bibinfo{year}{2009}), \eprint{0810.4110}.

\bibitem[{\citenamefont{Yin et~al.}(2009)\citenamefont{Yin, Yuan, Liu, Zhang,
  Bi, and Zhu}}]{Yin:2008bs}
\bibinfo{author}{\bibfnamefont{P.-f.} \bibnamefont{Yin}},
  \bibinfo{author}{\bibfnamefont{Q.}~\bibnamefont{Yuan}},
  \bibinfo{author}{\bibfnamefont{J.}~\bibnamefont{Liu}},
  \bibinfo{author}{\bibfnamefont{J.}~\bibnamefont{Zhang}},
  \bibinfo{author}{\bibfnamefont{X.-j.} \bibnamefont{Bi}}, \bibnamefont{and}
  \bibinfo{author}{\bibfnamefont{S.-h.} \bibnamefont{Zhu}},
  \bibinfo{journal}{Phys. Rev.} \textbf{\bibinfo{volume}{D79}},
  \bibinfo{pages}{023512} (\bibinfo{year}{2009}), \eprint{0811.0176}.

\bibitem[{\citenamefont{Bi et~al.}(2009)\citenamefont{Bi, Gu, Li, and
  Zhang}}]{Bi:2009md}
\bibinfo{author}{\bibfnamefont{X.-J.} \bibnamefont{Bi}},
  \bibinfo{author}{\bibfnamefont{P.-H.} \bibnamefont{Gu}},
  \bibinfo{author}{\bibfnamefont{T.}~\bibnamefont{Li}}, \bibnamefont{and}
  \bibinfo{author}{\bibfnamefont{X.}~\bibnamefont{Zhang}},
  \bibinfo{journal}{JHEP} \textbf{\bibinfo{volume}{04}}, \bibinfo{pages}{103}
  (\bibinfo{year}{2009}), \eprint{0901.0176}.

\bibitem[{\citenamefont{Cao et~al.}(2014)\citenamefont{Cao, Chen, and
  Gong}}]{Cao:2014cda}
\bibinfo{author}{\bibfnamefont{Q.-H.} \bibnamefont{Cao}},
  \bibinfo{author}{\bibfnamefont{C.-R.} \bibnamefont{Chen}}, \bibnamefont{and}
  \bibinfo{author}{\bibfnamefont{T.}~\bibnamefont{Gong}}
  (\bibinfo{year}{2014}), \eprint{1409.7317}.

\bibitem[{\citenamefont{Hagiwara et~al.}(1987)\citenamefont{Hagiwara, Peccei,
  Zeppenfeld, and Hikasa}}]{Hagiwara:1986vm}
\bibinfo{author}{\bibfnamefont{K.}~\bibnamefont{Hagiwara}},
  \bibinfo{author}{\bibfnamefont{R.~D.} \bibnamefont{Peccei}},
  \bibinfo{author}{\bibfnamefont{D.}~\bibnamefont{Zeppenfeld}},
  \bibnamefont{and} \bibinfo{author}{\bibfnamefont{K.}~\bibnamefont{Hikasa}},
  \bibinfo{journal}{Nucl. Phys.} \textbf{\bibinfo{volume}{B282}},
  \bibinfo{pages}{253} (\bibinfo{year}{1987}).

\bibitem[{\citenamefont{Falkowski and Mimouni}(2015)}]{Falkowski:2015krw}
\bibinfo{author}{\bibfnamefont{A.}~\bibnamefont{Falkowski}} \bibnamefont{and}
  \bibinfo{author}{\bibfnamefont{K.}~\bibnamefont{Mimouni}}
  (\bibinfo{year}{2015}), \eprint{1511.07434}.

\bibitem[{\citenamefont{Ellis and You}(2015)}]{Ellis:2015sca}
\bibinfo{author}{\bibfnamefont{J.}~\bibnamefont{Ellis}} \bibnamefont{and}
  \bibinfo{author}{\bibfnamefont{T.}~\bibnamefont{You}} (\bibinfo{year}{2015}),
  \eprint{1510.04561}.

\bibitem[{\citenamefont{Wells and Zhang}(2016)}]{Wells:2015eba}
\bibinfo{author}{\bibfnamefont{J.~D.} \bibnamefont{Wells}} \bibnamefont{and}
  \bibinfo{author}{\bibfnamefont{Z.}~\bibnamefont{Zhang}},
  \bibinfo{journal}{Phys. Rev.} \textbf{\bibinfo{volume}{D93}},
  \bibinfo{pages}{034001} (\bibinfo{year}{2016}), \bibinfo{note}{[Phys.
  Rev.D93,034001(2016)]}, \eprint{1507.01594}.

\bibitem[{\citenamefont{Bian et~al.}(2015)\citenamefont{Bian, Shu, and
  Zhang}}]{Bian:2015zha}
\bibinfo{author}{\bibfnamefont{L.}~\bibnamefont{Bian}},
  \bibinfo{author}{\bibfnamefont{J.}~\bibnamefont{Shu}}, \bibnamefont{and}
  \bibinfo{author}{\bibfnamefont{Y.}~\bibnamefont{Zhang}},
  \bibinfo{journal}{JHEP} \textbf{\bibinfo{volume}{09}}, \bibinfo{pages}{206}
  (\bibinfo{year}{2015}), \eprint{1507.02238}.

\bibitem[{\citenamefont{Cao et~al.}(2007)\citenamefont{Cao, Ma, Wudka, and
  Yuan}}]{Cao:2007fy}
\bibinfo{author}{\bibfnamefont{Q.-H.} \bibnamefont{Cao}},
  \bibinfo{author}{\bibfnamefont{E.}~\bibnamefont{Ma}},
  \bibinfo{author}{\bibfnamefont{J.}~\bibnamefont{Wudka}}, \bibnamefont{and}
  \bibinfo{author}{\bibfnamefont{C.~P.} \bibnamefont{Yuan}}
  (\bibinfo{year}{2007}), \eprint{0711.3881}.

\bibitem[{\citenamefont{Cao et~al.}(2011)\citenamefont{Cao, Chen, Li, and
  Zhang}}]{Cao:2009uw}
\bibinfo{author}{\bibfnamefont{Q.-H.} \bibnamefont{Cao}},
  \bibinfo{author}{\bibfnamefont{C.-R.} \bibnamefont{Chen}},
  \bibinfo{author}{\bibfnamefont{C.~S.} \bibnamefont{Li}}, \bibnamefont{and}
  \bibinfo{author}{\bibfnamefont{H.}~\bibnamefont{Zhang}},
  \bibinfo{journal}{JHEP} \textbf{\bibinfo{volume}{08}}, \bibinfo{pages}{018}
  (\bibinfo{year}{2011}), \eprint{0912.4511}.

\bibitem[{\citenamefont{Denner}(1993)}]{Denner:1991kt}
\bibinfo{author}{\bibfnamefont{A.}~\bibnamefont{Denner}},
  \bibinfo{journal}{Fortsch. Phys.} \textbf{\bibinfo{volume}{41}},
  \bibinfo{pages}{307} (\bibinfo{year}{1993}), \eprint{0709.1075}.

\bibitem[{\citenamefont{Aoki et~al.}(1982)\citenamefont{Aoki, Hioki, Konuma,
  Kawabe, and Muta}}]{Aoki:1982ed}
\bibinfo{author}{\bibfnamefont{K.~I.} \bibnamefont{Aoki}},
  \bibinfo{author}{\bibfnamefont{Z.}~\bibnamefont{Hioki}},
  \bibinfo{author}{\bibfnamefont{M.}~\bibnamefont{Konuma}},
  \bibinfo{author}{\bibfnamefont{R.}~\bibnamefont{Kawabe}}, \bibnamefont{and}
  \bibinfo{author}{\bibfnamefont{T.}~\bibnamefont{Muta}},
  \bibinfo{journal}{Prog. Theor. Phys. Suppl.} \textbf{\bibinfo{volume}{73}},
  \bibinfo{pages}{1} (\bibinfo{year}{1982}).

\bibitem[{\citenamefont{Passarino and Veltman}(1979)}]{Passarino:1978jh}
\bibinfo{author}{\bibfnamefont{G.}~\bibnamefont{Passarino}} \bibnamefont{and}
  \bibinfo{author}{\bibfnamefont{M.~J.~G.} \bibnamefont{Veltman}},
  \bibinfo{journal}{Nucl. Phys.} \textbf{\bibinfo{volume}{B160}},
  \bibinfo{pages}{151} (\bibinfo{year}{1979}).

\bibitem[{\citenamefont{'t~Hooft and Veltman}(1979)}]{tHooft:1978xw}
\bibinfo{author}{\bibfnamefont{G.}~\bibnamefont{'t~Hooft}} \bibnamefont{and}
  \bibinfo{author}{\bibfnamefont{M.~J.~G.} \bibnamefont{Veltman}},
  \bibinfo{journal}{Nucl. Phys.} \textbf{\bibinfo{volume}{B153}},
  \bibinfo{pages}{365} (\bibinfo{year}{1979}).

\bibitem[{\citenamefont{Stange and Willenbrock}(1993)}]{Stange:1993td}
\bibinfo{author}{\bibfnamefont{A.}~\bibnamefont{Stange}} \bibnamefont{and}
  \bibinfo{author}{\bibfnamefont{S.}~\bibnamefont{Willenbrock}},
  \bibinfo{journal}{Phys. Rev.} \textbf{\bibinfo{volume}{D48}},
  \bibinfo{pages}{2054} (\bibinfo{year}{1993}), \eprint{hep-ph/9302291}.

\bibitem[{\citenamefont{Cao et~al.}(2009{\natexlab{b}})\citenamefont{Cao, Chen,
  Larios, and Yuan}}]{PhysRevD.79.015004}
\bibinfo{author}{\bibfnamefont{Q.-H.} \bibnamefont{Cao}},
  \bibinfo{author}{\bibfnamefont{C.-R.} \bibnamefont{Chen}},
  \bibinfo{author}{\bibfnamefont{F.}~\bibnamefont{Larios}}, \bibnamefont{and}
  \bibinfo{author}{\bibfnamefont{C.-P.} \bibnamefont{Yuan}},
  \bibinfo{journal}{Phys. Rev. D} \textbf{\bibinfo{volume}{79}},
  \bibinfo{pages}{015004} (\bibinfo{year}{2009}{\natexlab{b}}).

\bibitem[{\citenamefont{Sirlin}(1980)}]{Sirlin:1980nh}
\bibinfo{author}{\bibfnamefont{A.}~\bibnamefont{Sirlin}},
  \bibinfo{journal}{Phys. Rev.} \textbf{\bibinfo{volume}{D22}},
  \bibinfo{pages}{971} (\bibinfo{year}{1980}).

\bibitem[{\citenamefont{Bardin et~al.}(1997)}]{Bardin:1997xq}
\bibinfo{author}{\bibfnamefont{D.~{\relax Yu}.} \bibnamefont{Bardin}}
  \bibnamefont{et~al.} (\bibinfo{year}{1997}), \eprint{hep-ph/9709229},
  \urlprefix\url{http://doc.cern.ch/cernrep/1995/95-03/95-03.html}.

\bibitem[{\citenamefont{Hahn et~al.}(2003)\citenamefont{Hahn, Hollik, Lorca,
  Riemann, and Werthenbach}}]{Hahn:2003ab}
\bibinfo{author}{\bibfnamefont{T.}~\bibnamefont{Hahn}},
  \bibinfo{author}{\bibfnamefont{W.}~\bibnamefont{Hollik}},
  \bibinfo{author}{\bibfnamefont{A.}~\bibnamefont{Lorca}},
  \bibinfo{author}{\bibfnamefont{T.}~\bibnamefont{Riemann}}, \bibnamefont{and}
  \bibinfo{author}{\bibfnamefont{A.}~\bibnamefont{Werthenbach}}
  (\bibinfo{year}{2003}), \eprint{hep-ph/0307132},
  \urlprefix\url{http://alice.cern.ch/format/showfull?sysnb=2382496}.

\bibitem[{\citenamefont{Mohr et~al.}(2015)\citenamefont{Mohr, Newell, and
  Taylor}}]{Mohr:2015ccw}
\bibinfo{author}{\bibfnamefont{P.~J.} \bibnamefont{Mohr}},
  \bibinfo{author}{\bibfnamefont{D.~B.} \bibnamefont{Newell}},
  \bibnamefont{and} \bibinfo{author}{\bibfnamefont{B.~N.} \bibnamefont{Taylor}}
  (\bibinfo{year}{2015}), \eprint{1507.07956}.

\bibitem[{\citenamefont{Hahn and Perez-Victoria}(1999)}]{Hahn:1998yk}
\bibinfo{author}{\bibfnamefont{T.}~\bibnamefont{Hahn}} \bibnamefont{and}
  \bibinfo{author}{\bibfnamefont{M.}~\bibnamefont{Perez-Victoria}},
  \bibinfo{journal}{Comput. Phys. Commun.} \textbf{\bibinfo{volume}{118}},
  \bibinfo{pages}{153} (\bibinfo{year}{1999}), \eprint{hep-ph/9807565}.

\bibitem[{\citenamefont{van Oldenborgh}(1991)}]{vanOldenborgh:1990yc}
\bibinfo{author}{\bibfnamefont{G.~J.} \bibnamefont{van Oldenborgh}},
  \bibinfo{journal}{Comput. Phys. Commun.} \textbf{\bibinfo{volume}{66}},
  \bibinfo{pages}{1} (\bibinfo{year}{1991}).

\bibitem[{\citenamefont{Khachatryan et~al.}(2015)}]{Khachatryan:2014rra}
\bibinfo{author}{\bibfnamefont{V.}~\bibnamefont{Khachatryan}}
  \bibnamefont{et~al.} (\bibinfo{collaboration}{CMS}), \bibinfo{journal}{Eur.
  Phys. J.} \textbf{\bibinfo{volume}{C75}}, \bibinfo{pages}{235}
  (\bibinfo{year}{2015}), \eprint{1408.3583}.

\bibitem[{\citenamefont{Aad et~al.}(2015{\natexlab{a}})}]{Aad:2015zva}
\bibinfo{author}{\bibfnamefont{G.}~\bibnamefont{Aad}} \bibnamefont{et~al.}
  (\bibinfo{collaboration}{ATLAS}), \bibinfo{journal}{Eur. Phys. J.}
  \textbf{\bibinfo{volume}{C75}}, \bibinfo{pages}{299}
  (\bibinfo{year}{2015}{\natexlab{a}}), \bibinfo{note}{[Erratum: Eur. Phys.
  J.C75,no.9,408(2015)]}, \eprint{1502.01518}.

\bibitem[{\citenamefont{Aad et~al.}(2015{\natexlab{b}})}]{Aad:2014tda}
\bibinfo{author}{\bibfnamefont{G.}~\bibnamefont{Aad}} \bibnamefont{et~al.}
  (\bibinfo{collaboration}{ATLAS}), \bibinfo{journal}{Phys. Rev.}
  \textbf{\bibinfo{volume}{D91}}, \bibinfo{pages}{012008}
  (\bibinfo{year}{2015}{\natexlab{b}}), \bibinfo{note}{[Erratum: Phys.
  Rev.D92,no.5,059903(2015)]}, \eprint{1411.1559}.

\bibitem[{\citenamefont{Khachatryan et~al.}(2016)}]{Khachatryan:2014rwa}
\bibinfo{author}{\bibfnamefont{V.}~\bibnamefont{Khachatryan}}
  \bibnamefont{et~al.} (\bibinfo{collaboration}{CMS}), \bibinfo{journal}{Phys.
  Lett.} \textbf{\bibinfo{volume}{B755}}, \bibinfo{pages}{102}
  (\bibinfo{year}{2016}), \eprint{1410.8812}.

\bibitem[{\citenamefont{Alwall et~al.}(2011)\citenamefont{Alwall, Herquet,
  Maltoni, Mattelaer, and Stelzer}}]{Alwall:2011uj}
\bibinfo{author}{\bibfnamefont{J.}~\bibnamefont{Alwall}},
  \bibinfo{author}{\bibfnamefont{M.}~\bibnamefont{Herquet}},
  \bibinfo{author}{\bibfnamefont{F.}~\bibnamefont{Maltoni}},
  \bibinfo{author}{\bibfnamefont{O.}~\bibnamefont{Mattelaer}},
  \bibnamefont{and} \bibinfo{author}{\bibfnamefont{T.}~\bibnamefont{Stelzer}},
  \bibinfo{journal}{JHEP} \textbf{\bibinfo{volume}{06}}, \bibinfo{pages}{128}
  (\bibinfo{year}{2011}), \eprint{1106.0522}.

\bibitem[{\citenamefont{Alloul et~al.}(2014)\citenamefont{Alloul, Christensen,
  Degrande, Duhr, and Fuks}}]{Alloul:2013bka}
\bibinfo{author}{\bibfnamefont{A.}~\bibnamefont{Alloul}},
  \bibinfo{author}{\bibfnamefont{N.~D.} \bibnamefont{Christensen}},
  \bibinfo{author}{\bibfnamefont{C.}~\bibnamefont{Degrande}},
  \bibinfo{author}{\bibfnamefont{C.}~\bibnamefont{Duhr}}, \bibnamefont{and}
  \bibinfo{author}{\bibfnamefont{B.}~\bibnamefont{Fuks}},
  \bibinfo{journal}{Comput. Phys. Commun.} \textbf{\bibinfo{volume}{185}},
  \bibinfo{pages}{2250} (\bibinfo{year}{2014}), \eprint{1310.1921}.

\bibitem[{\citenamefont{Ade et~al.}(2015)}]{Ade:2015xua}
\bibinfo{author}{\bibfnamefont{P.~A.~R.} \bibnamefont{Ade}}
  \bibnamefont{et~al.} (\bibinfo{collaboration}{Planck})
  (\bibinfo{year}{2015}), \eprint{1502.01589}.

\bibitem[{\citenamefont{Burgess et~al.}(2001)\citenamefont{Burgess, Pospelov,
  and ter Veldhuis}}]{Burgess:2000yq}
\bibinfo{author}{\bibfnamefont{C.~P.} \bibnamefont{Burgess}},
  \bibinfo{author}{\bibfnamefont{M.}~\bibnamefont{Pospelov}}, \bibnamefont{and}
  \bibinfo{author}{\bibfnamefont{T.}~\bibnamefont{ter Veldhuis}},
  \bibinfo{journal}{Nucl. Phys.} \textbf{\bibinfo{volume}{B619}},
  \bibinfo{pages}{709} (\bibinfo{year}{2001}), \eprint{hep-ph/0011335}.

\bibitem[{\citenamefont{Cirelli et~al.}(2006)\citenamefont{Cirelli, Fornengo,
  and Strumia}}]{Cirelli:2005uq}
\bibinfo{author}{\bibfnamefont{M.}~\bibnamefont{Cirelli}},
  \bibinfo{author}{\bibfnamefont{N.}~\bibnamefont{Fornengo}}, \bibnamefont{and}
  \bibinfo{author}{\bibfnamefont{A.}~\bibnamefont{Strumia}},
  \bibinfo{journal}{Nucl. Phys.} \textbf{\bibinfo{volume}{B753}},
  \bibinfo{pages}{178} (\bibinfo{year}{2006}), \eprint{hep-ph/0512090}.

\bibitem[{\citenamefont{Lopez~Honorez et~al.}(2007)\citenamefont{Lopez~Honorez,
  Nezri, Oliver, and Tytgat}}]{LopezHonorez:2006gr}
\bibinfo{author}{\bibfnamefont{L.}~\bibnamefont{Lopez~Honorez}},
  \bibinfo{author}{\bibfnamefont{E.}~\bibnamefont{Nezri}},
  \bibinfo{author}{\bibfnamefont{J.~F.} \bibnamefont{Oliver}},
  \bibnamefont{and} \bibinfo{author}{\bibfnamefont{M.~H.~G.}
  \bibnamefont{Tytgat}}, \bibinfo{journal}{JCAP}
  \textbf{\bibinfo{volume}{0702}}, \bibinfo{pages}{028} (\bibinfo{year}{2007}),
  \eprint{hep-ph/0612275}.

\bibitem[{\citenamefont{Hambye et~al.}(2009)\citenamefont{Hambye, Ling,
  Lopez~Honorez, and Rocher}}]{Hambye:2009pw}
\bibinfo{author}{\bibfnamefont{T.}~\bibnamefont{Hambye}},
  \bibinfo{author}{\bibfnamefont{F.~S.} \bibnamefont{Ling}},
  \bibinfo{author}{\bibfnamefont{L.}~\bibnamefont{Lopez~Honorez}},
  \bibnamefont{and} \bibinfo{author}{\bibfnamefont{J.}~\bibnamefont{Rocher}},
  \bibinfo{journal}{JHEP} \textbf{\bibinfo{volume}{07}}, \bibinfo{pages}{090}
  (\bibinfo{year}{2009}), \bibinfo{note}{[Erratum: JHEP05,066(2010)]},
  \eprint{0903.4010}.

\end{thebibliography}

\end{document}